\documentclass[longauth]{aa}

\usepackage{booktabs}
\usepackage{orcidlink}
\usepackage[varg]{txfonts}
\usepackage{enumitem}
\usepackage{upgreek}
\begin{document} 

   \title{Refined parameters, formation, and dynamical stability of the wild exoplanet system K2-312$\equiv$HD\,80653}
   \titlerunning{The wild exoplanet system K2-312}
   \author{L.~Naponiello\orcidlink{0000-0001-9390-0988}\inst{1}\fnmsep\thanks{Corresponding author: luca.naponiello@inaf.it}
   \and E.~Poretti\orcidlink{0000-0003-1200-0473}\inst{2,3}
   \and K.~Rice\orcidlink{0000-0002-6379-9185}\inst{4,5}
   \and A.~S.~Bonomo\orcidlink{0000-0002-6177-198X}\inst{1}
   \and L.~Malavolta\orcidlink{0000-0002-6492-2085}\inst{6}
   \and M.~Stalport\orcidlink{0000-0003-0996-6402}\inst{7,8}
   \and A.~Vanderburg\orcidlink{0000-0001-7246-5438}\inst{9}
   \and C.~Ziegler\orcidlink{0000-0002-0619-7639]}\inst{10}
   \and L.~Affer\orcidlink{0000-0001-5600-3778}\inst{11}
   \and M.~Cecconi\inst{3}
   \and A.~Collier~Cameron\orcidlink{0000-0002-8863-7828}\inst{12}
   \and R.~Cosentino\orcidlink{0000-0003-1784-1431}\inst{3}
   \and M.~Damasso\orcidlink{0000-0001-9984-4278}\inst{1}
   \and X.~Dumusque\orcidlink{0000-0002-9332-2011}\inst{13}
   \and Y.~N.~E.~Eschen\orcidlink{0009-0006-6397-2503}\inst{14}
   \and A.~Ghedina\orcidlink{0000-0003-4702-5152}\inst{3}
   \and D.~W.~Latham\orcidlink{0000-0001-9911-7388}\inst{9}
   \and M.~L\'opez-Morales\orcidlink{0000-0003-3204-8183}\inst{15}
   \and T.~Lu\orcidlink{0000-0003-0834-8645}\inst{16}   
   \and A.~Massa\inst{17}
   \and A.~Mortier\orcidlink{0000-0001-7254-4363}\inst{18}
   \and B.~A.~Nicholson\orcidlink{0000-0003-1360-4404}\inst{19,20}
   \and L.~Palethorpe\orcidlink{0000-0002-1664-4105}\inst{21}
   \and F.~A.~Pepe\orcidlink{0000-0002-9815-773X}\inst{13}
   \and A.~Sozzetti\orcidlink{0000-0002-7504-365X}\inst{1}
   \and S.~Udry\orcidlink{0000-0001-7576-6236}\inst{13}
   \and T.~G.~Wilson\orcidlink{0000-0001-8749-1962}\inst{14}
   }
   \authorrunning{L.~Naponiello et al.}

    \institute{INAF - Osservatorio Astrofisico di Torino, Via Osservatorio 20, 10025 Pino Torinese, Italy
    \and %2
    INAF - Osservatorio Astronomico di Brera, via E. Bianchi 46, 23807 Merate (LC)
    \and %3
    Fundación Galileo Galilei-INAF, Rambla José Ana Fernandez Pérez 7, 38712 Breña Baja, TF, Spain
    \and %4
    Centre for Planetary Sciences, The University of Edinburgh, Edinburgh, EH9 3FF, UK
    \and %5
    SUPA, Institute for Astronomy, University of Edinburgh, The Royal Observatory, Blackford Hill, Edinburgh, EH9 3HJ, UK
    \and %6
    Dipartimento di Fisica e Astronomia “Galileo Galilei”, Universit\`a di Padova, Vicolo del l’Osservatorio 3, I-35122 Padova, Italy
    \and %7
    Space sciences, technologies and astrophysics research Institute, Université de Liège, Allée du 6 Août 19C, 4000 Liège, Belgium
    \and %8
    Astrobiology Research Unit, Université de Liège, Allée du 6 Août 19C, B-4000 Liège, Belgium
    \and %9
    Center for Astrophysics | Harvard \& Smithsonian, 60 Garden Street, Cambridge, MA 02138, USA
    \and %10
    Department of Physics, Engineering and Astronomy, Stephen F. Austin State University, 1936 North St, Nacogdoches, TX 75962, USA
    \and %11
    Istituto Nazionale di Astrofisica, Osservatorio Astronomico di Palermo, Palermo, Italy
    \and %12
    Centre for Exoplanet Science, School of Physics and Astronomy, University of St Andrews, St Andrews, UK
    \and %13
    Observatoire de Genève, Département d’Astronomie, Université de Genève, Chemin Pegasi 51, 1290 Versoix, Switzerland
    \and %14
    Department of Physics, University of Warwick, Gibbet Hill Road, Coventry CV4 7AL, UK
    \and %15
    Space Telescope Science Institute, 3700 San Martin Drive, Baltimore MD 21218, USA
    \and %16
    Center for Computational Astrophysics, Flatiron Institute, 162 5th Avenue, New York, NY 10010, USA
    \and %17
    Dipartimento di Fisica, Università degli Studi di Torino, via Pietro Giuria 1, 10125 Torino, Italy
    \and %18
    School of Physics \& Astronomy, University of Birmingham, Edgbaston, Birmingham, B15 2TT, UK
    \and %19
    Centre for Astrophysics, University of Southern Queensland, Toowoomba, Australia
    \and %20
    Sub-department of Astrophysics, University of Oxford, Oxford, UK
    \and %21
    HH Wills Physics Laboratory, University of Bristol, Tyndall Avenue, Bristol, BS8 1TL, UK
    %\and %22
    }

   \date{Received on June 3rd, 2026; Accepted on July 16th, 2026.}

% \abstract{}{}{}{}{} 
% 5 {} token are mandatory
 
 \abstract
   {The architecture of planetary systems hosting ultra-short-period (USP) planets is a key diagnostic for understanding formation and migration scenarios. The presence of outer giant companions in these systems is of particular interest to test theories regarding dynamical effects and pebble accretion.}
   {We present an extended radial velocity (RV) monitoring of the bright star K2-312$\equiv$HD\,80653, known to host a rocky USP super-Earth ($P_b=0.720$\,d). Previous studies identified a long-term trend and subsequently a Keplerian signal due to an outer highly eccentric giant planet, K2-312\,c. We aim to refine the orbital parameters of K2-312\,c by precisely monitoring its periastron passage and to model the formation and dynamical evolution of the system.}
   {We analyzed a set of 237 HARPS-N high-resolution spectra, extending the observation baseline of previous literature by almost 4 years. We performed a joint analysis of the RVs together with K2 and TESS photometry to refine the ephemerides and properties of the two planets. To account for stellar activity, we coupled the Keplerian models with a Gaussian processes regression.}
   {K2-312\,c is a cold Jupiter on a wide orbit (orbital period refined to $P_c=871.32$\,d), with a minimum mass of $M_c \sin i \sim 5\,M_{\rm Jup}$ and a refined eccentricity of $e_c \sim 0.85$. It is among the most eccentric cold Jupiters known in multi-planet systems, and the only one that is highly eccentric and has a USP planet companion. Our simulations suggest that planet-planet scattering between two giant planets could have driven K2-312\,c to its current high eccentricity, ejected the other giant, and still allowed for the survival of K2-312\,b. The extended observation baseline further allowed us to identify the stellar rotation period and a long activity cycle, while a new K2 reduction improved the significance of the secondary eclipse detection for K2-312\,b.}
   {}


   \maketitle
   \nolinenumbers
%
%________________________________________________________________

\section{Introduction}\label{sec:intro}
Exoplanetary systems hosting planets in extreme conditions represent crucial case studies for theoretical models of planet formation and evolution. The system K2-312$\equiv$EPIC 251279430$\equiv$HD\,80653 consists of a G0V star ($V=9.45$\,mag) and two very different planets, both with extreme properties: 

\begin{itemize}

    \item a transiting ultra-short-period (USP; $P_b=0.72$\,d; semimajor axis $a=0.017$\,AU) rocky super-Earth ($R_b=1.6\,\rm R_{\oplus}$; $\rho_b=7.4$\,g\,cm$^{-3}$), K2-312\,b$\equiv$HD\,80653\,b. This planet was characterized by \citet[hereafter F20]{Frustagli2020} using K2 photometry and high-precision radial velocities (RVs) collected with the High Accuracy Radial velocity Planet Searcher in the Northern hemisphere (HARPS-N) spectrograph at the Telescopio Nazionale Galileo \citep{Cosentino2012} from November 2018 to May 2019, and
~\\
    \item a long-period ($P_c=921.2\pm10.8\,$d; $a\sim1.96\pm0.03$\,AU) and  massive ($M_c \sin{i}\sim5\,\rm M_{\rm Jup}$) gaseous giant planet, K2-312\,c$\equiv$HD\,80653\,c, on a highly eccentric ($e\sim0.85$) orbit, as first characterized by \citet[hereafter B23]{Bonomo2023} through an extension of the RV monitoring from November 2019 to June 2020, intended to unveil the origin of the long-term trend first noted by F20.

   \end{itemize}

The USP small planets ($R_p < 4\,\rm R_\oplus$) are rare. They are found around $\sim 0.5\%$ of solar-like stars \citep{sanchis-ojeda2014}. Although their exact origin remains a subject of debate, they are generally thought not to form in situ, but rather to migrate to their current extreme orbits either through disk-driven migration \citep[e.g.,][]{Pu2019} or via multi-body dynamical interactions followed by tidal circularization \citep[e.g.,][]{Winn2018, Petrovich2019}. K2-312\,b is even rarer among USP small planets because (i) it has one of the largest radii \citep{sanchis-ojeda2014} and (ii) it does not have detected short-period siblings, unlike the majority of systems with USP small planets \citep{Winn2018, Zhu2025}. The giant planet K2-312\,c is also quite unusual. It is one of 13 known cold Jupiters (CJs; i.e., gaseous giant planets with $1 < a < 10$\,au and $M_p > 0.3\,\rm M_{\rm Jup}$) on highly eccentric ($e > 0.8$) orbits. 

K2-312 is part of the small sample of $\sim 15$ systems known to host transiting inner ($P< 100$\,d) small planets and CJs \citep{Bonomo2025, BryanLee2025}. The occurrence rate of CJs in small planet systems is $\sim 11-13\%$ at an average stellar metallicity and mass, depending on the adopted mass range for the CJs \citep{Bonomo2025}. 
In particular, it is one of three known systems with a USP small planet and a CJ (the others are Kepler-407 \citep{Weiss2024} and WASP-47 \citep{Almenara2016, Vanderburg2017}), and it is the only system with a highly eccentric CJ that coexists with an inner small planet. This makes K2-312 a sort of wild system composed of two very peculiar exoplanets. Nonetheless, it is not surprising that the host star has a supersolar metallicity ([Fe/H]>0.1\,dex) and mass ($M_\star>1\,\rm M_\odot$). In this range, small planets and CJs might be correlated \citep{Bonomo2025, BryanLee2025}.

We present an updated analysis of the K2-312 system (Sect.\,\ref{sec:analysis}) using a further extended dataset of HARPS-N RV observations collected from December 2024 to May 2025 (Sect.\,\ref{sec:spectroscopy}). In particular, the new RV data allowed us to fully cover the periastron passage of the outer planet K2-312\,c, refine its orbital parameters and mass, and confirm its remarkable eccentricity (Sect.\,\ref{sec:solution}). The improved extraction of the K2 light curve allowed us to confirm the parameters of K2-312\,b and increase the significance of its secondary eclipse while simultaneously modeling its phase variations (Sect.\,\ref{sec:eclipse}). We also improved the K2-312\,b ephemeris with the new photometry (Sect.\,\ref{sec:photometry}) from the Transiting Exoplanet Survey Satellite (TESS; \citealt{Ricker2014}). Finally, we carried out simulations of planet formation and migration to reproduce the architecture of this wild system and to understand its origin better (Sect.\,\ref{sec:formation}).

\section{Host star properties}\label{sec:host}
K2-312 is a metal-rich ([Fe/H]=$0.255 \pm 0.065$), G ($T_{\rm eff}=5959\pm61$~K) dwarf (F20). Since the Gaia DR3 parallax was not available at the time of the discovery paper, we recomputed the stellar parameters, namely mass, radius, and age, with a differential evolution Markov chain Monte Carlo Bayesian framework through the tool {\tt EXOFASTv2} (\citealt{2017ascl.soft10003E, Eastman2019}; see also \citealt{Naponiello2025} for more details). To this end, we simultaneously modeled the MIST evolutionary tracks (e.g., \citealt{Paxton2015}) and the stellar spectral energy distribution (SED) using the Tycho-2 $B_{\rm T}$ and $V_{\rm T}$, APASS Johnson $B$, the 2MASS $J$, $H$, and $K_{\rm s}$ magnitudes, and WISE $W1$, $W2$, and $W3$ magnitudes (Table~\ref{tab:star}).
We imposed a Gaussian prior on the Gaia DR3 parallax and on the effective temperature $T_{\rm eff}$ and [Fe/H] derived in F20. 
We found $M_\star=1.150^{+0.063}_{-0.069}~\rm M_\odot$, $R_\star=1.199 \pm 0.033~\rm R_\odot$, and age $t=3.3^{+2.7}_{-2.0}$~Gyr, in excellent agreement with the values determined by F20, with all parameters consistent within $1.0\sigma$. On the basis of these parameters, we can classify K2-312 as a G0V star \citep{PecautMamajek2013}. The best fit of the stellar SED is shown in Fig.~\ref{fig:stellarSED}. 

%%% TABLE 1 - Stellar Parameters %%%%%%%%%%%%%%%%%%%%%%%%%%%%
\begin{table}
\centering %
\caption{Stellar parameters of HD\,80653.}\label{tab:star} %
\resizebox{1.02\hsize}{!}{
\begin{tabular}{lccc}
\hline %
\hline  \\[-8pt]%%
 & Unit & Value & Source \\
\hline  \\[-6pt]%%
%\multicolumn{1}{l}{\large{Identifiers}} \\ [2pt] %
HD \dotfill & \dotfill & 80653 & HD \\
K2 \dotfill & \dotfill & 312 & Kepler \\
EPIC \dotfill & \dotfill & 251279430 & K2 \\
%HIP \dotfill & \dotfill& ... & HIP \\
TOI \dotfill & \dotfill & 5556 & TOI catalog \\
%TIC \dotfill & \dotfill & 151759246 & TIC \\
Tycho-2 \dotfill & \dotfill & 825-1097-1 & Tycho-2 \\
2MASS \dotfill & \dotfill & {\tiny J09212142+1422046} & 2MASS \\
{\it Gaia}  \dotfill & \dotfill & {\tiny 606477252238780160} & {\it Gaia}~DR3 \\ [6pt] %
%\hline \\[-6pt]%
%\multicolumn{1}{l}{\large{Astrometric properties}} \\ [2pt] %
$\alpha$\,(J2016.0) \dotfill & h & \,\,\,09:21:21.38 & {\it Gaia}~DR3 \\
$\delta$\,(J2016.0) \dotfill & deg & $+$14:22:04.33 & {\it Gaia}~DR3 \\
$\pi$ \dotfill & mas & $9.260 \pm 0.022$ & {\it Gaia}~DR3 \\
$\mu_\alpha \cos{\delta}$ \dotfill & mas/yr & $-33.297\pm 0.023$ & {\it Gaia}~DR3 \\
$\mu_\delta$ \dotfill & mas/yr & $-14.040\pm 0.017$\,\,\,\, & {\it Gaia}~DR3 \\
$d$ \dotfill & pc & $107.99\pm 0.25$ & This work$^1$ \\ [6pt] %
%\hline \\[-6pt]%
%\multicolumn{1}{l}{\large{Photometric properties}} \\ [2pt] %
$B_{\rm T}$ \dotfill & mag & $10.269\pm 0.031$\,\,\, & Tycho-2 \\ 
$B$ \dotfill & mag & $10.107\pm 0.021$\,\,\, & APASS \\
$V_{\rm T}$ \dotfill & mag & $9.528\pm 0.023$ & Tycho-2 \\ 
$G$ \dotfill & mag & $9.3139\pm 0.0028$ & {\it Gaia}~DR3 \\
$J$ \dotfill & mag & $8.315\pm 0.023$ & 2MASS \\
$H$ \dotfill & mag & $8.079\pm 0.029$ & 2MASS \\
$K_{\rm S}$ \dotfill  & mag & $8.018\pm 0.021$ & 2MASS \\ 
%$i'$ \dotfill & mag & $10...\pm0...$ & APASS \\
$W1$ \dotfill & mag & $7.959\pm 0.024$ & AllWISE \\
$W2$ \dotfill & mag & $8.004\pm 0.020$ & AllWISE \\     
$W3$ \dotfill & mag & $8.011\pm 0.021$ & AllWISE \\ 
%$W4$ \dotfill & mag & $xx.xx\pm xx.xx$ & AllWISE \\ 
$A_V$ \dotfill & mag & $<0.065$ & This work$^1$ \\ [6pt] %   
%\hline \\[-6pt]%
%\multicolumn{1}{l}{\large{Stellar parameters}} \\ [2pt] %
Spectral type \dotfill & & G0\,V & This work$^2$ \\
$L_{\star}$ \dotfill & $L_{\sun}$ & $1.628\pm 0.071$ & This work$^1$ \\
$M_{\star}$ \dotfill & $M_{\sun}$ & $1.150^{+0.063}_{-0.069}$ & This work$^1$ \\
$R_{\star}$ \dotfill & $R_{\sun}$ & $1.199\pm 0.033$ & This work$^1$ \\
$T_{\rm eff}$ \dotfill & K & $5959\pm 61$ & F20, B23\\ 
%$v\sin{i_{\star}}$ \dotfill & km\,s$^{-1}$ & $xx.xx\pm xx.xx$ & This work$^4$ \\
%$\log g_{\star}$ \dotfill & cgs & $xx.xx\pm xx.xx$ & This work$^1$ \\
${\rm [Fe/H]}$ \dotfill & dex & $0.255\pm0.065$ & F20, B23 \\ 
$\log g_{\star}$ \dotfill & cgs & $4.342^{+0.032}_{-0.037}$ & This work$^1$ \\
$v\sin{i_{\star}}$ \dotfill & km\,s$^{-1}$ & $3.50\pm 0.50$ & F20 \\
%$\xi$ \dotfill & km\,s$^{-1}$ & $xx.xx\pm xx.xx$ & This work$^2$ \\
%${\rm [Mg/H]}$ \dotfill & dex & $xx.xx\pm xx.xx$ & This work$^2$ \\
%${\rm [Si/H]}$ \dotfill & dex & $xx.xx\pm xx.xx$ & This work$^2$ \\
$\rho_{\star}$\dotfill & g\,cm$^{-3}$ & $0.941\pm 0.095$ & This work$^1$ \\
%$\upsilon \sin{i_{\star}}$\dotfill & km\,s$^{-1}$ & $1..\pm0..$ & This work \\
%$\log A{\rm (Li)_{\rm NLTE}}$\dotfill &  & $1...\pm0...$ & This work \\
%$P_{\rm rot}$\dotfill & d & $29.9\pm3.5$ & This work \\
%$\log R^{\prime}_{\rm HK}$\dotfill & dex & $-4.885\pm0.007$ & This work$^2$ \\
% old    Age $\tau_{\rm \star}$ & $[\mathrm{Gyr}]$ & $2.65$ & $\:\pm\:$ & $0.31$ & Sec.\:\ref{sec:star_prop}
Age\dotfill & Gyr & $3.3^{+2.7}_{-2.0}$ & This work$^1$ \\
%$U^{(a)}$ \dotfill & km\,s$^{-1}$ & $xx.x$ & This work \\ 
%$V^{(a)}$ \dotfill & km\,s$^{-1}$ & $x.x$ & This work \\ 
%$W^{(a)}$ \dotfill & km\,s$^{-1}$ & $x.x$ & This work \\ 
\hline %
\end{tabular}
}
\tablebib{TESS Primary Mission TOI catalogue \citep{Guerrero2021}; HD \citep{Cannon1924}; Tycho-2 \citep{hog}; 2MASS \citep{Cutri2003}; {\it Gaia} DR3 \citep{Gaia2023}; APASS \citep{Henden2016}; AllWISE \citep{Cutri2013}.}
\begin{flushleft}
\footnotemark[1]{\small From the {\tt EXOFASTv2} modelling (this work).} \\
\footnotemark[2]{\small According to the \citet{PecautMamajek2013} calibration [v. 2022].} \\
\end{flushleft}
\end{table}

\section{Observations and data reduction}\label{sec:obs}

\subsection{Photometry}\label{sec:photometry}

The star K2-312 was observed continuously for about 80 days during Campaign 16 (between December 2017 and February 2018) of the K2 mission \citep{Howell2014}, the extended mission of the Kepler spacecraft, with an integration time of 1800\,sec (long-cadence mode). We produced our own light curve using a modified version of the \texttt{K2SFF} pipeline \citep[see Sect.\,\ref{sec:eclipse}]{Vanderburg2014, Vanderburg2016ApJS}.

The target was subsequently observed by TESS during Sectors\,45-46 (November-December 2021), 72 (November 2023) and 1751 (a short 7-day sector obtained in January 2026 that was interrupted by observations of the comet 3I/ATLAS). We used the light curve generated using the Presearch Data Conditioning Simple Aperture Photometry (PDC-SAP; \citealt{Stumpe2012, Stumpe2014}, \citealt{Smith2012}), which is provided by the TESS SPOC pipeline and retrieved via the Python package \texttt{lightkurve} \citep{lightkurve} from the Mikulski Archive for Space Telescopes (MAST). The PDC-SAP photometry is already corrected for dilution from other objects contained within the aperture using the module called Create Optimal Apertures \citep{Bryson2010, Bryson2020}, although in this case, the TIC contamination ratio is $<0.01$. Since short-cadence TESS data are available, we adopted the 120\,s exposure time for Sectors\,45-46 (SPOC reduction) and the 200\,s exposure time for Sector\,72 (TESS-SPOC), while Sector 1751 was not used due to its short duration and the lack of a SPOC reduction at this stage.

To account for residual photometric variability, we modeled the correlated noise component using a Gaussian process (GP) with a Matérn kernel, as implemented in the \texttt{juliet}\footnote{\url{https://github.com/nespinoza/juliet/}} package \citep{Espinoza2019}. Independent sets of hyperparameters were adopted for each dataset (K2, TESS 120\,s, and TESS 200\,s). Since the fits with two independent terms yielded consistent results, we used a common activity component for the consecutive TESS Sectors 45 and 46.

\subsection{HARPS-N spectroscopy}\label{sec:spectroscopy}

We collected a total of 237 high-resolution spectra of K2-312 using HARPS-N. The observations spanned a time baseline of 6.5 years, from November 2018 to May 2025. This dataset includes the 114 spectra used in F20, 97 additional spectra used in B23, and 22 new spectra acquired specifically to constrain the orbit of planet c by catching its periastron passage (see Table\,\ref{tab:dataset}).

The data were reduced using the latest version of the HARPS-N data reduction software (DRS v. 3.3.12; see, e.g., \citealt{Dumusque2021}), extracting RVs and activity indicators via the cross-correlation function (CCF) method with a G2 mask. We made use of the DACE platform\footnote{\url{https://dace.unige.ch}}, which also provides additional spectroscopic activity diagnostics, including the Ca II H\&K S-index \citep{Vaughan1978}.

\subsection{SOAR imaging}\label{sec:highres}

High-angular resolution imaging is needed to search for nearby sources that can contaminate transits, resulting in an underestimated planetary radius, or be the source of astrophysical false positives, such as background eclipsing binaries. We searched for stellar companions to K2-312 (TOI-5556) with speckle imaging on the 4.1 m Southern Astrophysical Research (SOAR) telescope \citep{Tokovinin2018}, observing in the Cousins I-band, a similar visible band-pass as TESS. More details on observations within the SOAR TESS survey are available in \citealt{Ziegler2020}. The $5\sigma$ detection sensitivity and speckle autocorrelation functions from the observations are shown in Fig.\,\ref{fig:highres}. No nearby stars were detected within $3\arcsec$ of K2-312.

\section{Analysis}\label{sec:analysis}

\subsection{Two-planet model}\label{sec:simple_model}

We first modeled the RV data using a sum of Keplerian orbits, including a circular orbit for the USP planet b, with Gaussian priors centered on the period and time of transit derived by F20, and one orbit for planet c, with wide uniform priors. In particular, for the giant, the eccentricity was parameterized in terms of $\sqrt{e}\sin{\omega}$ and $\sqrt{e}\cos{\omega}$, both assigned uniform priors $\mathcal{U}(-1,+1)$, the orbital period was assigned the prior $\mathcal{U}(700,1200)$\,d, while the RV semi-amplitude was sampled in $\mathcal{U}(0,300)$\,m\,s$^{-1}$. An uncorrelated jitter term ($\sigma_{\mathrm{jit}}$) was then added to the model to account for any residual white noise. We employed the dynamic nested sampling algorithm via the \texttt{dynesty} package \citep{Speagle2020}, as included in the Python wrapper \texttt{juliet}, to sample the parameter space. This approach allowed us to simultaneously derive the posterior distributions of the planetary parameters and compute the Bayesian log-evidence ($\ln \mathcal{Z}$) for a robust model comparison.

This simple two-planet model, however, failed to account for a long-term trend that can be seen by eye (Fig.\,\ref{fig:full_fit}) as a RV difference ($\sim15$\,m\,s$^{-1}$) between the data obtained after the first (at $\mathrm{BJD}-2457000\sim2000$) and third periastron passages ($\mathrm{BJD}-2457000\sim3700$). Moreover, the generalized Lomb-Scargle (GLS) periodogram of the RV residuals shows a $<0.1$\% false-alarm probability (FAP) peak at $\sim17$\,d (Fig.~\ref{fig:GLS_res}) that is related to the stellar rotation period (see below and Fig.\,5 in F20). This peak was detected with and without modeling the long-term trend. 

Fig.~\ref{fig:trend} shows the RV residuals (top panel) and the corresponding activity indicators (lower panels): the CCF area (FWHM $\times$ contrast; \citealt{Cameron2019}), the bisector span (BIS; see, e.g., \citealt{Queloz2001}), and the Mount Wilson S-index. Fig.~\ref{fig:GLS} shows the GLS periodogram of the same datasets. The long-term RV trend is evident in all the activity indices, unresolved from 0.0~d$^{-1}$ with the current dataset. The RV residuals and S-index exhibit a modest Pearson correlation coefficient of $0.44$. In particular, after correcting the S-index for this trend, another significant ($<0.1$\% FAP) peak appears at $\sim17$\,d, suggesting that stellar activity contaminated the RV dataset to some extent. The presence of such signals motivated the inclusion of a GP component in the RV fit.

\subsection{GP model}\label{sec:activity}

To properly take the rotation period (short) and stellar activity cycle (long) into account, we employed a GP kernel inspired by \citet{Basilicata2024} that was the sum of a quasi-periodic (QP) kernel, as defined by \citealt{Rajpaul2015}, and a squared exponential kernel,

\begin{eqnarray}
k(t,t')=H^2_{\mathrm{rot}}\,e^{-\frac{\sin^2[\pi(t-t')/P_{\mathrm{rot}}]}{2O^2_{\mathrm{amp}}}-\frac{(t-t')^2}{2P^2_{\mathrm{dec}}}} + H^2_{\mathrm{cycle}}\,e^{-\frac{(t-t')^2}{2\lambda^2_{\mathrm{cycle}}}} ,
\end{eqnarray}

where $(t,t')$ are the epochs at two different RV observations, $H_{\mathrm{cycle}}$ and $\lambda_{\mathrm{cycle}}$ are the amplitude and the timescale of the squared exponential kernel, and $H_{\mathrm{rot}}$, $O_{\mathrm{amp}}$, and $P_{\mathrm{dec}}$ are the amplitude, the length scale, and the evolutionary timescale of the QP kernel, with $P_{rot}$ being its characteristic period. Finally, we added the formal RV uncertainty at time $t$ and the uncorrelated jitter term ($\sigma_{\mathrm{jit}}$) in quadrature.

This new model properly accounts for the activity signals, as the new RV residuals show no significant trends. Compared to a model that only uses the QP kernel, the Bayesian evidence is not significantly higher ($\ln \Delta\mathcal{Z}|^{\mathrm{QP+Exp}}_{\mathrm{QP}} \sim 2$), and we further note that a quadratic trend in addition to the QP kernel can describe the data equally well ($\ln \Delta\mathcal{Z}|^{\mathrm{QP+Exp}}_{\mathrm{QP+trend}} \sim 1$); however, we adopted the model with the combined kernels because it provided a more realistic representation of the long-term trend likely due to a stellar activity cycle.

\subsection{Orbital solution}\label{sec:solution}

\begin{figure}
    \centering
    \includegraphics[width=1\linewidth]{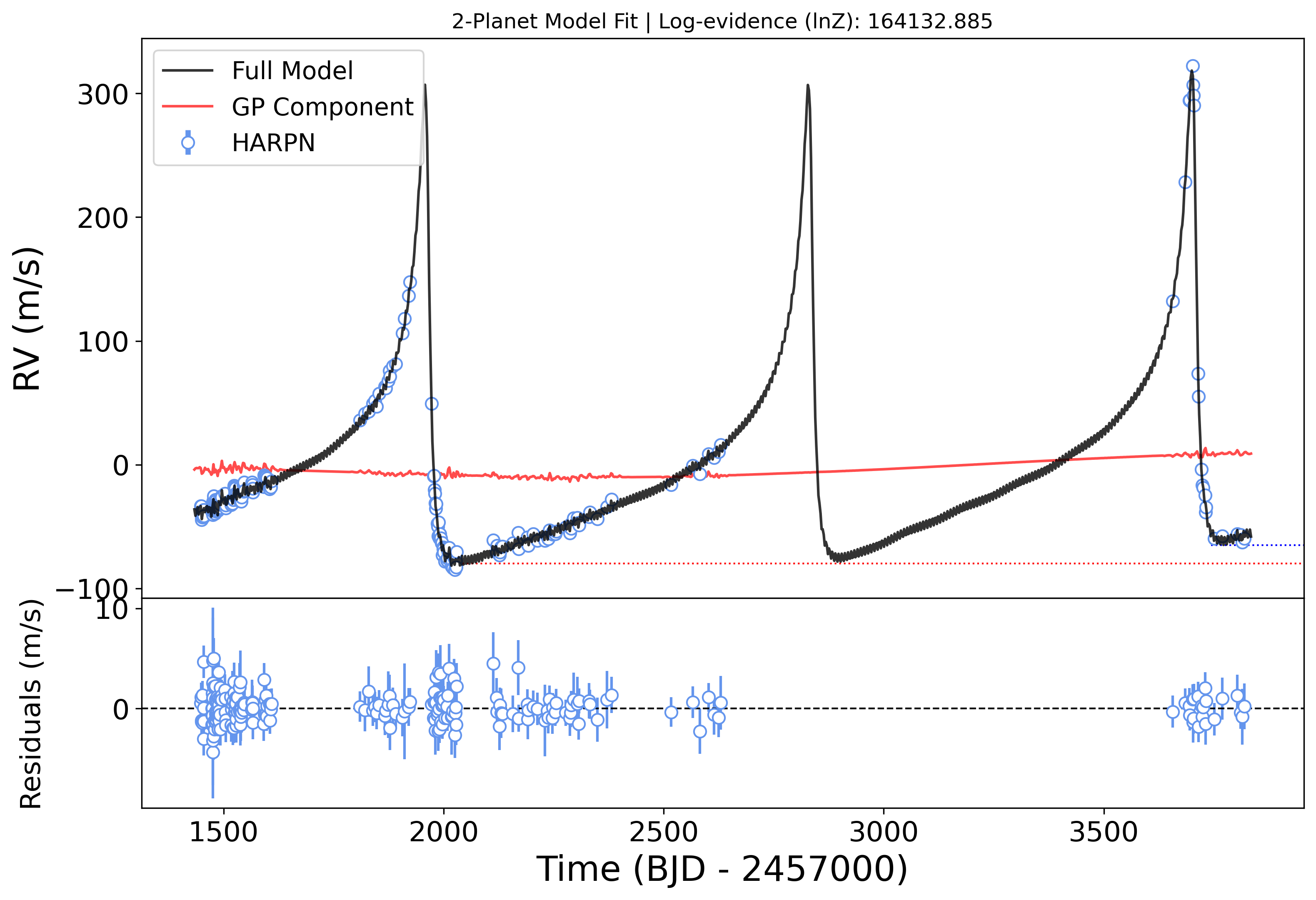}
    \caption{HARPS-N RV measurements of K2-312 in blue. The best model fit is displayed in black in the top panel, along with its GP component in red. The residuals of the RVs are shown in the bottom panel. The dotted red and blue lines represent the bottom RV level after the first and third periastron passage, respectively. The dataset used in F20 and B23 ends at 1600 and 2650 BJD$-2457000$, respectively.}
    \label{fig:full_fit}
\end{figure}

The best-fit solution from a joint transit+RV model confirmed the highly eccentric nature of planet c. We derived a period of $P_c=871.30\pm0.13$\,d, an eccentricity $e_c=0.8435\pm0.0013$, and an RV semi-amplitude $K_c=191.6\pm1.3$\,m\,s$^{-1}$ (Fig.\,\ref{fig:full_fit}). Notably, the newly derived period differs from the one reported in B23 by $4.5\sigma$. This discrepancy likely arises from (i) the partial coverage of the first periastron passage of K2-312\,c and (ii) the long-term trend due to activity, which was not evident in the data analyzed by B23 and thus was not modeled at the time. According to the new ephemeris, the predicted transit of K2-312\,c could have occurred right at the beginning of the K2 observations, with the nominal mid-transit time approximately 6\,h after the start of the time series. However, the uncertainty on the predicted mid-transit time is about 9\,h, so an earlier transit before the start of the observations cannot be ruled out.

For planet b, our analysis yields a substantially improved determination of the orbital period, ($P = 0.71957926^{+0.00000042}_{-0.00000036}$\,d), reducing the uncertainty by a factor of $\sim 50$ with respect to the previous literature value ($0.719573 \pm 0.000021$\,d). The new period is fully consistent with previous measurements and provides a significantly more precise ephemeris for future observations.

\begin{figure}
    \centering
    \includegraphics[width=1\linewidth]{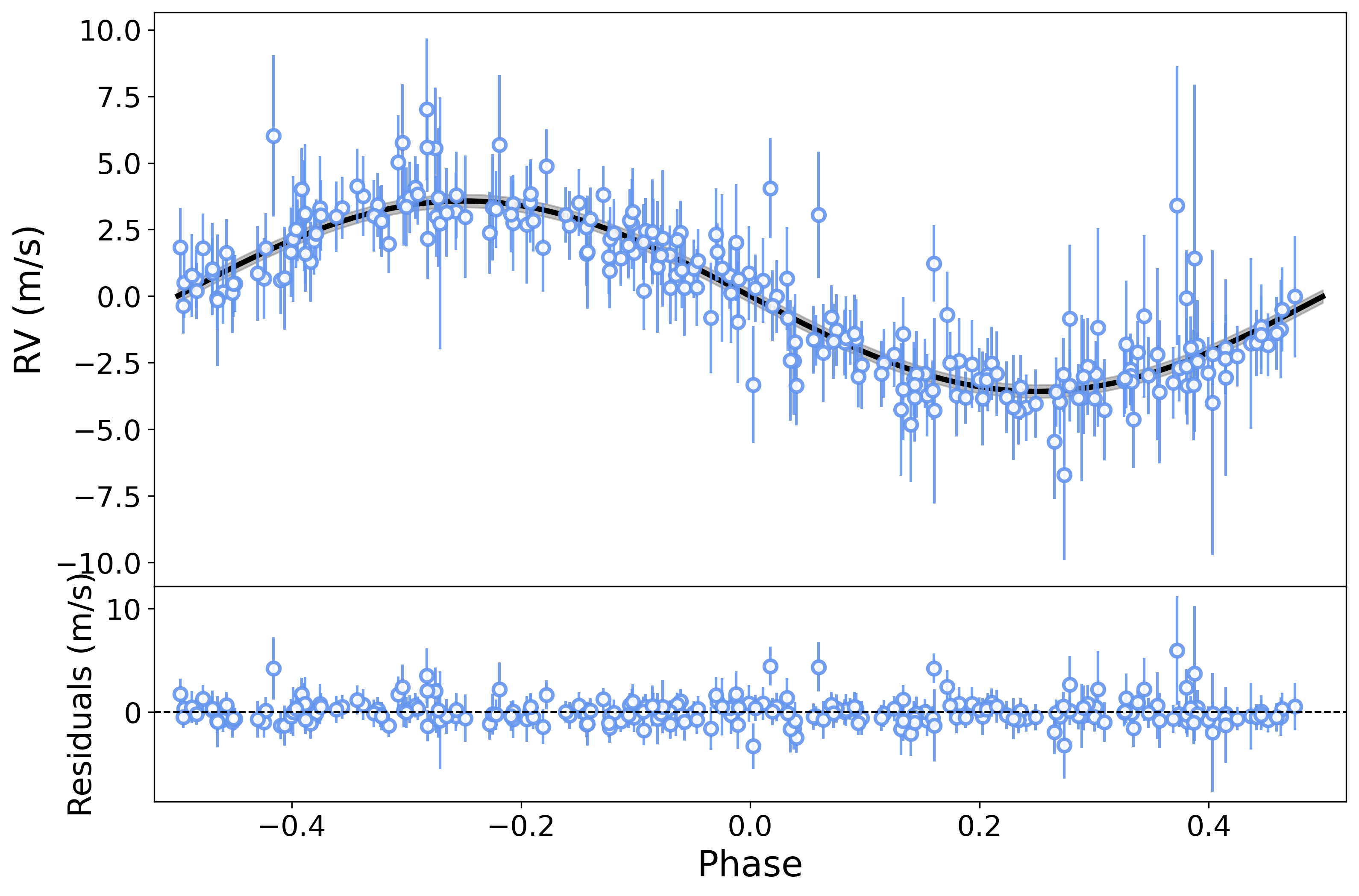}
    \includegraphics[width=1\linewidth]{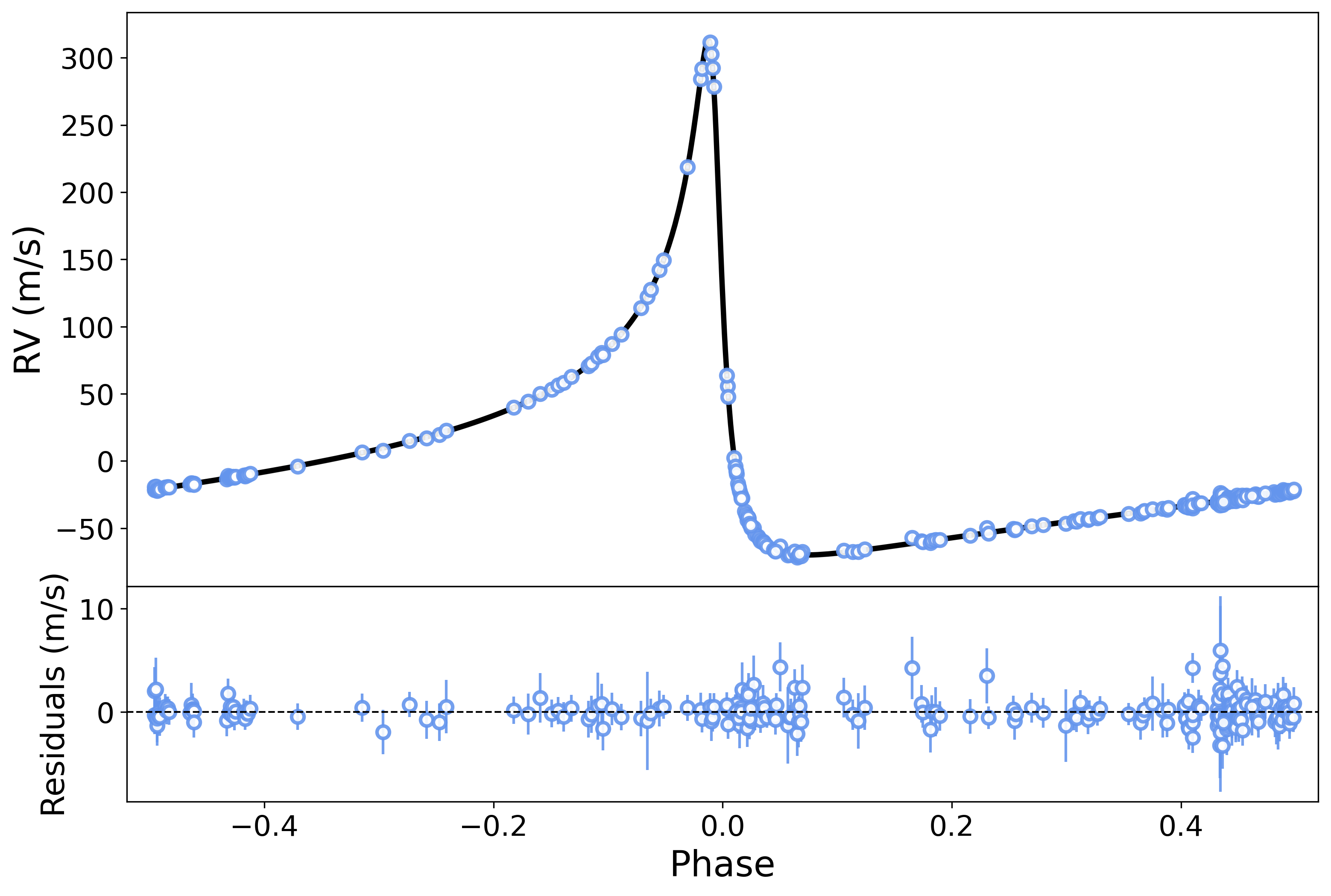}
    \caption{Phase-folded HARPS-N RVs to the period of planet b (top) and c (bottom) after removing the activity signals model, along with their best-fit models and residuals.}
    \label{fig:phased_plot}
\end{figure}

\begin{figure*}
    \centering
    \includegraphics[width=0.49\linewidth]{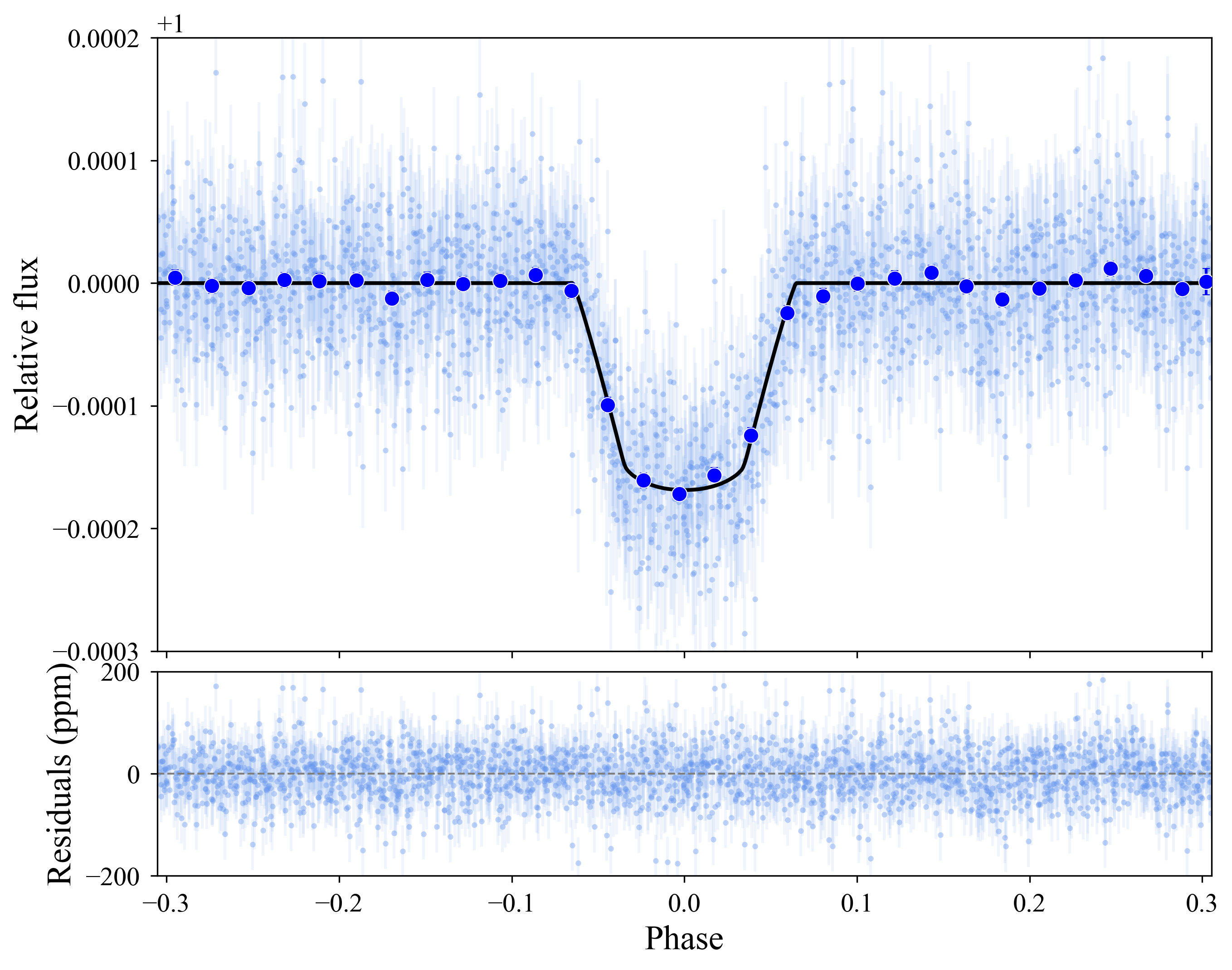}
    \includegraphics[width=0.49\linewidth]{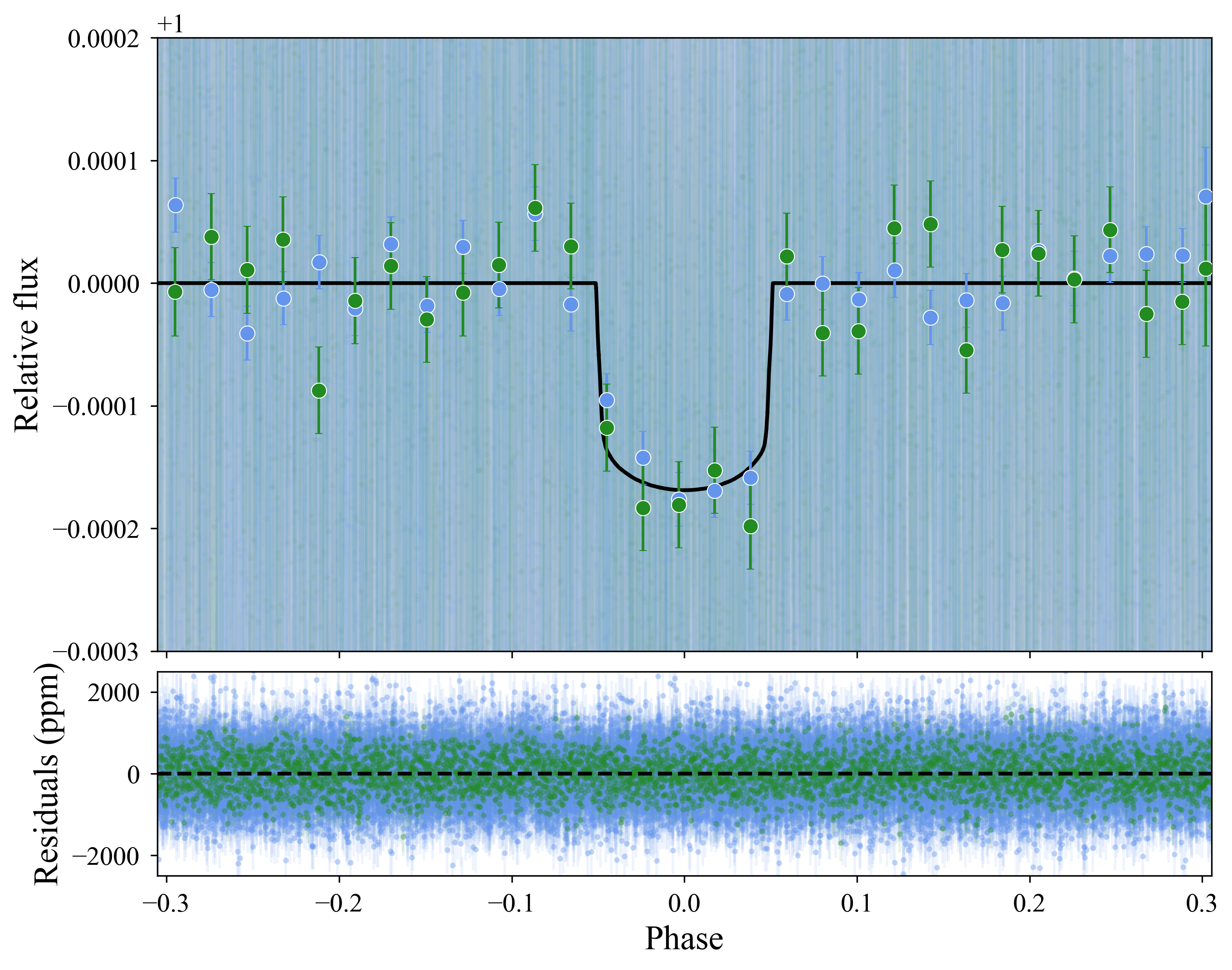}
    \caption{Phase-folded Kepler (left) and TESS (right) light curves of K2-312\,b, along with the best-fit model in black. TESS data points from sectors with 120 and 200 sec exposures are represented in green and cyan. The Kepler cadence is 1800 seconds. The large circles show phase bins of 0.02. A zoomed-out version of the TESS light curve is presented in Fig.\,\ref{fig:zoom}.}
    \label{fig:transit}
\end{figure*}

Moreover, from the GP hyper-parameters, we recovered a stellar rotation period of $P_{rot}=18.98^{+0.66}_{-0.74}$\,d, which is slightly longer than the peak in the RV residuals of the two-planet model and the peak in the S-index, but well compatible with the rotation period estimated in F20 and B23 ($19.08\pm0.5$ and $19.55^{+0.57}_{-0.50}$\,d, respectively). Fig.\,\ref{fig:phased_plot} shows the Keplerian signals due to the planets, and Fig.\,\ref{fig:transit} shows the K2 and TESS phase-folded transits of K2-312\,b.

\begin{table}
\centering
\caption{Final parameters for the planets.}\label{tab:pparameters}
\renewcommand{\arraystretch}{1.2}
\resizebox{0.85\hsize}{!}{\begin{tabular}{lcc}
    \hline\hline
     Parameters & K2-312\,b & K2-312\,c \\
     %& (1p model) & (1p model) \\ [2pt]
\hline \\[-6pt]%
\multicolumn{1}{l}{\large{Transit and orbital}} \\[2pt]
$K$ (m\,s$^{-1}$)\dotfill & $3.66\pm0.20$ & $191.6\pm1.3$ \\
$P_{\rm orb}$ (d)\dotfill & $0.71957926^{+0.00000042}_{-0.00000036}$ & $871.32\pm0.13$ \\
$T_{\rm 0}$ (BJD - $2\,458\,000$)\dotfill & $95.56741^{+0.00051}_{-0.00052}$ & $968.72\pm0.23$ \\
$T_{\rm 14}$ (h)\dotfill & $1.725^{+0.029}_{-0.028}$ & -- \\
$R_{\rm p}/R_{\star}$\dotfill & $0.01242^{+0.00022}_{-0.00029}$ & -- \\ 
$b$\dotfill & $0.429^{+0.075}_{-0.107}$ & -- \\ %\equiv a \cos{i}(1-e^2)/(R_{\star}(1+e\sin{\omega}))
$i$ (deg)\dotfill & $81.7^{+2.2}_{-1.8}$ & -- \\ 
$a/R_{\star}$\dotfill & $2.973^{+0.096}_{-0.100}$ & $338\pm11$ \\
$q_1$ (K2)\dotfill & $0.17^{+0.20}_{-0.10}$ & -- \\
$q_2$ (K2)\dotfill & $0.38^{+0.37}_{-0.26}$ & -- \\
$q_1$ (TESS)\dotfill & $0.15^{+0.26}_{-0.11}$ & -- \\
$q_2$ (TESS)\dotfill & $0.34^{+0.38}_{-0.25}$ & -- \\
$\sqrt{e}\sin\omega_p$\dotfill & -- & $0.6060\pm0.0056$ \\
$\sqrt{e}\cos\omega_p$\dotfill & -- & $0.6901^{+0.0050}_{-0.0049}$ \\[2pt]
%\multicolumn{1}{l}{Limb-darkening coefficients} \\
%$u_{1}$\dotfill & ~~~~~~~~~~~~~~~~~~~~~~~~~~~~~~~~~~~~~~~~~~$0.27^{+0.18}_{-0.13}$ \\ [2pt]
%$u_{2}$\dotfill & ~~~~~~~~~~~~~~~~~~~~~~~~~~~~~~~~~~~~~~~~~~$0.35^{+0.17}_{-0.16}$ \\ [2pt]
%\hline \\[-6pt]%
%\multicolumn{1}{l}{{RV parameters}} \\
%$\gamma$ (m\,s$^{-1}$)\dotfill & ?? & ?? \\
%RV jitter HARPS-N (m\,s$^{-1}$)\dotfill & ~~~~~~~~~~~~~~~~~~~~~~~~~~~~~~~~~~~~~~~~~~$2.93^{+0.35}_{-0.32}$ \\ [2pt]
%RV jitter SOPHIE  (m\,s$^{-1}$)\dotfill & $24^{+18}_{-12}$ \\ [2pt]
%\hline \\[-6pt]%
\multicolumn{1}{l}{\large{Derived}} \\[2pt]
$M_{\rm p}$\dotfill & $5.61^{+0.38}_{-0.37}\,M_{\oplus}$ & $5.31\pm0.21\,M_{\mathrm{Jup}}$ \\
%$M_{\rm p}\sin{i}$ ($M_{\oplus}$)\dotfill & ... & ... \\[2pt]
$R_{\rm p}$ ($R_{\oplus}$)\dotfill & $1.621^{+0.054}_{-0.057}$ & -- \\
$\rho_{\rm p}$ (g\,cm$^{-3}$)\dotfill & $7.24^{+0.97}_{-0.81}$ & -- \\ [2pt]
$\log{g_{\rm p}}$ (cgs)\dotfill & $1.32^{+0.32}_{-0.28}$ & -- \\
$a$ (au)\dotfill & $0.01657^{+0.00070}_{-0.00071}$ & $1.882^{+0.080}_{-0.081}$ \\ 
$T_{\rm eq}^{(\dagger)}$ (K)\dotfill & $2438^{+47}_{-45}$ & $229^{+5}_{-4}$ \\
$e$\dotfill & $0$ & $0.8435\pm0.0013$ \\ [2pt]
$\omega_p$ (deg)\dotfill & 90 & $41.3\pm0.5$ \\ 
TSM\dotfill & $35.4^{+3.6}_{-3.5}$ & -- \\ [2pt]
$u_1$ (K2)\dotfill & $0.30^{+0.24}_{-0.20}$ & -- \\
$u_2$ (K2)\dotfill & $0.09^{+0.30}_{-0.24}$ & -- \\
$u_1$ (TESS)\dotfill & $0.22^{+0.25}_{-0.16}$ & -- \\
$u_2$ (TESS)\dotfill & $0.04^{+0.25}_{-0.17}$ & -- \\
%$u_1$ (LCO$_Y$)\dotfill & $0.72^{+0.30}_{-0.28}$ & $0.78^{+0.30}_{-0.27}$ \\
%$u_2$ (LCO$_Y$)\dotfill & $0.06^{+0.28}_{-0.27}$ & $-0.02^{+0.26}_{-0.26}$ \\
%$u_1$ (LCO$_{zs}$)\dotfill & $0.70^{+0.28}_{-0.25}$ & $0.79^{+0.28}_{-0.25}$ \\
%$u_2$ (LCO$_{zs}$)\dotfill & $0.00^{+0.23}_{-0.24}$ & $-0.11^{+0.23}_{-0.24}$ \\
\multicolumn{1}{l}{\large{Instrumental}} \\[2pt]
$\sigma_{\textsf{K2}}$ (ppm)\dotfill & \multicolumn{2}{c}{$34.8\pm1.5$} \\
$\sigma_{\textsf{TESS}_{120}}$ (ppm)\dotfill & \multicolumn{2}{c}{$341.6\pm6.5$} \\
$\sigma_{\textsf{TESS}_{200}}$ (ppm)\dotfill & \multicolumn{2}{c}{$6.5^{+47}_{-6}$} \\
$\sigma_{\textsf{K2,GP}}$ (ppm)\dotfill & \multicolumn{2}{c}{$592\pm25$} \\
$\rho_{\textsf{K2,GP}}$ (d)\dotfill & \multicolumn{2}{c}{$0.998^{+0.001}_{-0.003}$} \\
$\sigma_{\textsf{TESS,GP}_{120}}$ (ppm)\dotfill & \multicolumn{2}{c}{$69\pm9$} \\
$\rho_{\textsf{TESS,GP}_{120}}$ (d)\dotfill & \multicolumn{2}{c}{$0.42^{+0.15}_{-0.11}$} \\
$\sigma_{\textsf{TESS,GP}_{200}}$ (ppm)\dotfill & \multicolumn{2}{c}{$185^{+0.31}_{-0.24}$} \\
$\rho_{\textsf{TESS,GP}_{200}}$ (d)\dotfill & \multicolumn{2}{c}{$0.62^{+0.14}_{-0.11}$} \\
%$\overline{\mu}_{\textsf{HARPS}}$ (m\,s$^{-1}$)\dotfill & $-xx.21^{+0.65}_{-0.61}$ & -- \\
$\overline{\mu}_{\textsf{HARPS-N}}$ (m\,s$^{-1}$)\dotfill & \multicolumn{2}{c}{$8330\pm4$} \\
%$\sigma_{\textsf{w,HARPS}}$ (m\,s$^{-1}$)\dotfill & $2.70^{+0.61}_{-0.52}$ & -- \\
$\sigma_{\textsf{w,HARPS-N}}$ (m\,s$^{-1}$)\dotfill & \multicolumn{2}{c}{$0.80^{+0.24}_{-0.19}$} \\
$H_{\mathrm{cycle}}$ (m\,s$^{-1}$)\dotfill & \multicolumn{2}{c}{$9.6^{+4.9}_{-3.1}$} \\
$\lambda_{\mathrm{cycle}}$ (d)\dotfill & \multicolumn{2}{c}{$788^{+312}_{-196}$} \\
$H_{\mathrm{rot}}$ (m\,s$^{-1}$)\dotfill & \multicolumn{2}{c}{$3.0^{+0.29}_{-0.26}$} \\
$O_{\mathrm{amp}}$ \dotfill & \multicolumn{2}{c}{$0.062^{+0.019}_{-0.009}$} \\
$P_{\mathrm{rot}}$ (d)\dotfill & \multicolumn{2}{c}{$18.98^{+0.66}_{-0.74}$} \\
$P_{\mathrm{dec}}$ (d)\dotfill & \multicolumn{2}{c}{$13.78^{+3.11}_{-2.73}$} \\

%$\Theta$\dotfill & ?? & -- \\
%$\log_{10}{\langle F \rangle}$ (cgs) \dotfill & ?? & ?? \\ [2pt]
    \bottomrule
\end{tabular}
}
\tablefoot{Best-fit median values, with upper and lower 68\% credibility bands as errors, as extracted from the posterior distribution of the relative models. $^{(\dagger)}$ This is the equilibrium temperature for a zero Bond albedo and uniform heat redistribution to the night side. The eccentricity upper limit on K2-312\,c is constrained at the confidence level of 1$\sigma$.}
\end{table}

\subsection{Eclipse fitting}\label{sec:eclipse}

A tentative detection of the secondary eclipse of K2-312\,b was reported by F20 and \citet{Singh2022} at a $2.2\sigma$ confidence level, prompting us to revisit its detection with dedicated care. Unfortunately, the secondary eclipse is too shallow to be detected by TESS, so we attempted to improve on existing reductions of the K2 light curve. F20 and \citet{Singh2022} both used light curves produced by the K2SFF pipeline. \citet{Singh2022} used a standard version of the light curve (as downloaded from the MAST), which was produced by de-correlating the raw measured flux against the spacecraft pointing drift. The standard version of the K2SFF pipeline does not incorporate any advanced knowledge of the astrophysical signals in the dataset, and it can therefore sometimes distort or partially suppress real astrophysical signals such as transits or secondary eclipses. To avoid this complication, it is common to refine the K2SFF systematics correction by performing a simultaneous nonlinear least-squares optimization of the systematics correction along with the transit models and stellar variability \citep{Vanderburg2016ApJS}. This was the strategy taken by F20, who used a refined version of the K2SFF light curve that modeled the primary transits of K2-312\,b simultaneously with the systematics corrections and the stellar variability. However, the statistical significance of the secondary eclipse was comparable to the result by \citet{Singh2022}. 

We therefore sought to further refine the K2SFF systematics correction. We used a similar strategy to \citet{Vanderburg2016ApJS} and F20 of performing a simultaneous least-squares fit, but we made two key improvements. First, we included a model component to describe the secondary eclipse of K2-312 b explicitly in our model, along with the primary transit. This ensured that the eclipse was not suppressed by the systematics correction. The other change we made was to somewhat increase the flexibility and complexity of the systematics model by increasing the number of free parameters used to model the flux variations due to spacecraft systematics as the Kepler pointing drifted. In particular, we found that increasing the number of free parameters beyond the standard value of 15 improved the correction of the K2 pointing systematics, while higher values did not produce noticeable improvements. We therefore adopted 25 as a reasonable compromise between flexibility and robustness. The large number of data points per segment (200-250) prevented the risk of overfitting. In most cases, 15 parameters are sufficient to model the spacecraft systematics, but we found for this saturated star that the variations were sufficiently complex for a 15-parameter model to leave clear systematic  residuals in the light curve. Increasing the complexity of our model improved the photometric precision of the light curve by about 50\% compared to the reductions used by F20 and \citet{Singh2022} and brought the light-curve precision more in line with the typical photometric precision attained by K2 for stars with a similar brightness.  

\begin{figure}
    \centering
    \includegraphics[width=1\linewidth]{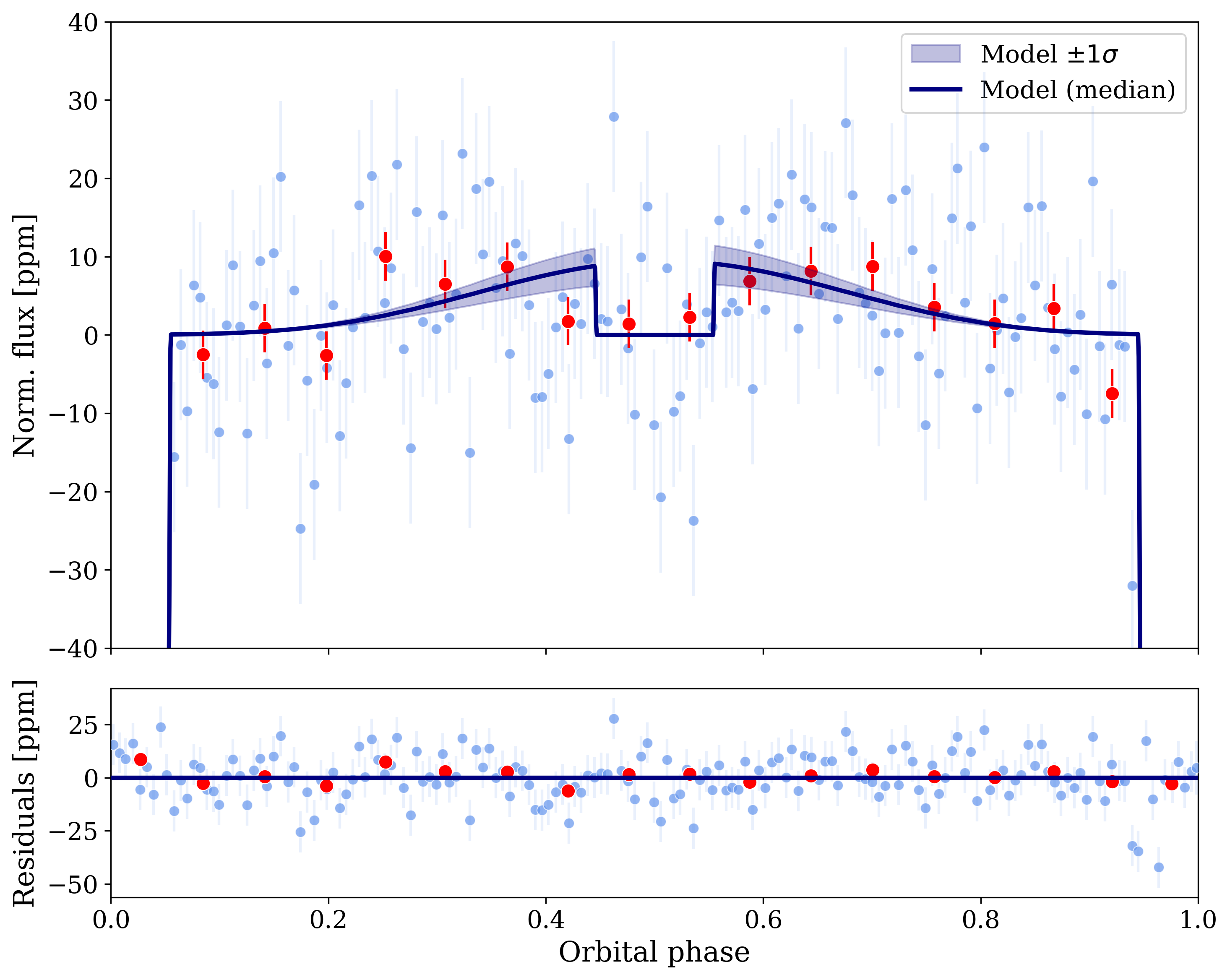}
    \caption{Phase-folded secondary eclipse of K2-312\,b as seen by Kepler. The cyan and red circles represent 20 and 185 K2 binned fluxes, respectively.}
    \label{fig:eclipse}
\end{figure}

We then modeled the secondary eclipse together with the primary transits adopting the physically motivated thermal phase-curve model  \citep{Morris2022} as implemented in the \texttt{kelp} package\footnote{\url{https://github.com/Jayshil/kelp}}. Instead of fitting the eclipse depth as a free parameter, \texttt{kelp} first constructs a dayside temperature map via spherical harmonics and subsequently integrates it over the \textit{Kepler} transmission function to generate the observed phase curve. The amplitude of the day-night temperature contrast is controlled by the $C^1_1$ spherical harmonic coefficient, which we fitted with a uniform prior $\mathcal{U}(0, 0.5)$; all remaining phase-curve parameters were held fixed at their standard values: hotspot offset $\Delta\phi = 0\degr$, dimensionless drag frequency $\omega_\mathrm{drag} = 4.5$, fluid number $\alpha = 0.6$, and greenhouse parameter $f' = 0.707$ \citep{Morris2022}. The systematic flux variability was modeled simultaneously with a GP with a \texttt{Matern-3/2} kernel as implemented in \texttt{george} \citep{Ambikasaran2015}, whose amplitude and length-scale  hyperparameters were fitted with log-uniform priors. The ephemerides ($P$, $T_0$) were fixed and a strict Gaussian prior on $R_\mathrm{p}/R_\star$ was imposed from the primary transit analysis (Sect.\,\ref{sec:solution}). We obtained $C^1_1=0.414^{+0.053}_{-0.074}$ and
\begin{equation}
    \delta_\mathrm{ecl} = 9.1^{+2.4}_{-2.7}\;\mathrm{ppm}
    \quad (\sim\!3.4\sigma),
    \label{eq:eclipse_depth}
\end{equation}
which represents a meaningful detection, consistent with but more significant than the literature value. The improvement is attributable primarily to our updated \texttt{K2SFF} reduction. The best-fit eclipse model is shown in Fig.\,\ref{fig:eclipse}, where the low-amplitude phase curve can also be appreciated.

Because the Kepler band-pass spans the optical range ($\sim$420--900\,nm), the observed eclipse depth is expected to include contributions from the dayside thermal emission and reflected stellar light. At the equilibrium temperature of K2-312\,b,
\begin{equation}
    T_\mathrm{eq} = T_{\star}\left(\frac{1}{4}\frac{R_{\star}}{a}\right)^{1/2}
    = 2438\,\mathrm{K}
    \quad (A_\mathrm{B}=0,\ \text{full redistribution}),
\end{equation}
the Planck function is still far from its peak at optical wavelengths; a
numerical integration of $B_\lambda(T_\mathrm{eq})$ over the Kepler response function yields a thermal contribution of only $\sim 0.5$\,ppm,
which is negligible compared to the measured depth. The geometric albedo directly follows from
\begin{equation}
    A_g = \frac{\delta_\mathrm{ecl}}{(R_p/a)^2}
        = \frac{\delta_\mathrm{ecl}}{(R_p/R_\star)^2/(a/R_\star)^2},
\end{equation}
where $(R_p/a)^2 = 17.4$\,ppm is the maximum reflected-light
depth attainable at $A_g=1$. Substituting the measured depth, we obtain
\begin{equation}
    A_g = 0.52 \pm 0.15,
\end{equation}
where the uncertainty accounts for the propagated contributions from
$\delta_\mathrm{ecl}$, $R_p/R_\star$, and $a/R_\star$. The 3$\sigma$ upper limit is $A_g < 0.97$, which is formally consistent with any
albedo value.

For completeness, we also derived the dayside brightness temperature under the conservative assumption that the entire eclipse signal originates from thermal emission, treating reflected light as negligible. This yielded
\begin{equation}
    T_\mathrm{bright} = 3440^{+130}_{-160}\,\mathrm{K}
    \quad (<\!3790\,\mathrm{K}\ \text{at}\ 3\sigma),
\end{equation}
obtained by numerically inverting the ratio of band-integrated Planck functions. This value should be regarded as a firm upper limit on the true dayside temperature: since reflected light almost certainly contributes to the observed depth, the actual thermal emission must be lower. A confirmation of the eclipse at longer wavelengths (e.g.,\ \textit{JWST}/MIRI), where thermal emission dominates and the reflected-light contribution is negligible, is required to break this degeneracy and constrain $A_g$ and $T_\mathrm{bright}$ independently. This is essential for distinguishing between a highly reflective dayside, an exposed global magma ocean, or a localized rock-vapor atmosphere.

\subsection{Alternative analysis}\label{sec:alt}

We performed an independent analysis of the photometric and spectroscopic data with the Bayesian framework \texttt{PyORBIT}\footnote{\url{https://pyorbit.readthedocs.io/}} \citep{Malavolta2016, Malavolta2018}, which was used in the discovery paper (F20) with the goal of testing the robustness of our results under different assumptions. We simultaneously modeled K2 and TESS photometry, the HARPS-N radial velocities, and four activity indices (FWHM and BIS of the CCF, Mount Wilson S-index, and H$\alpha$) as extracted from the HARPS-N DRS. We assumed a circular orbit for the transiting planet and left the eccentricity and argument of pericenter free to vary for the nontransiting planet(s), using the parameterization of \cite{Eastman2013}. 

The modeling of K2 data included the planetary signal, a second-order  polynomial trend to normalize the light curve and take care of the instrumental trend, and a Gaussian process to model short-term stellar activity. The planetary model of the USP planet included planetary transits and eclipses, modeled using \texttt{BATMAN} \citep{Kreidberg2015}, and a phase-curve model consisting of a sinusoid. The model accounted for the binning effect due to the 30-minute exposure of K2 photometry \citep{Kipping2010}. We did not include other effects, such as nightside emission, phase offset, or light travel-time effects, because the precision and duration of the K2 campaign are insufficient to detect or characterize them. We employed a GP with the exponential-sine periodic (ESP) kernel implemented in the \texttt{S+LEAF}\footnote{\url{https://gitlab.unige.ch/delisle/spleaf}} package \citep{Delisle2020, Delisle2022}. This kernel represents a fast approximation of the quasi-periodic kernel \citep{Rajpaul2015}, when a sufficient number of harmonics are included (four, in our case). We initially forced the kernel period and scale to match the rotational period and timescale decay hyperparameters constrained by the spectroscopic activity indexes, but the results were suboptimal. Specifically, the photometry drove the value of the rotational period of the star toward a value inconsistent with the spectroscopic-derived value and the $v \sin i$ by F20. We then decided to use an independent set of hyperparameters to model the activity in the K2 light curve. For this kernel, we imposed a Gaussian prior on the coherence scale of $0.350 \pm 0.035$ following \cite{Perger2021}, the condition on the hyperparameters suggested by \cite{Rajpaul2021} to ensure that the kernel has at least one nontrivial turning point, namely $P_\mathrm{dec}^2 > \frac{3}{2\pi} P_\mathrm{rot}^2\,O_\mathrm{amp}^2$, and forced the decay scale to be at least twice the periodic scale.

The lower precision of the TESS light curve prevented the detection of the phase-curve signal, so we filtered instrumental and stellar signals with the Biweight filter implemented in \texttt{wotan}\footnote{\url{https://github.com/hippke/wotan}} \citep{Hippke2019} after masking the USP transit signals and with a filtering window of 1.0~d, and modeled only the transit signal of the USP. This choice was mostly driven by the complexity of the model because we also included a multidimensional GP for the spectroscopic dataset. For the K2 and TESS data, we used independent limb-darkening coefficients with the parameterization by \cite{Kipping2013} for the quadratic law and uniform non-informative priors, and independent jitter parameters, while all the planetary and stellar parameters (planetary radius, impact parameter, time of inferior conjunction, and stellar density) were in common.

For the RV modeling, we employed a multivariate Gaussian process \citep{Rajpaul2015, Nardiello2022, Mantovan2024} with the \texttt{S+LEAF} ESP kernel, using the same priors as in the photometric case. Jointly with the RV, we modeled the  activity indexes above to constrain the underlying GP model and its first derivative (only for the RV and BIS data). Each dataset was modeled with a second-order polynomial to account for the long-term variation trend observed in the activity indexes, with independent offset and jitter parameters. We applied the same condition on the hyperparameters as in the analysis of the K2 data. The RV signals of the planets were included in the modeling of the RV dataset, with the orbital period and RV semi-amplitude explored in natural space for the transiting planet and in logarithmic space for the non-transiting planets. 

We explored the parameter space following two approaches. The first approach envisaged a global exploration of the parameter space with the differential evolution code \texttt{pyDE}\footnote{\url{https://github.com/hpparvi/PyDE}} (\citealt{Parviainen2015}, $10^5$ generations) followed by a Markov chain Monte Carlo through \texttt{emcee}\footnote{\url{https://github.com/dfm/emcee}} \citep{ForemanMackey2013} with $10^5$ steps, a burn-in of $3\times10^4$ steps, and 220 walkers. In the second approach, we used the nested sampling code \texttt{dynesty}\footnote{\url{https://github.com/joshspeagle/dynesty}} \citep{Speagle2020, Koposov2025} with 1000 live points. The posteriors from the two methods were perfectly super-imposable, although the MCMC analysis produced confidence intervals with slightly larger error bars. As a conservative approach, we report the results of the MCMC analysis here.

Our analysis resulted in a RV semi-amplitude of $K_b=3.61\pm0.22$ m~s$^{-1}$ and a scaled planetary radius of $R_p/R_s=0.01229_{-0.00039}^{+0.00029}$ for the USP planet, corresponding to a mass of $M_b=5.59\pm0.41\,M_{\oplus}$ and  $R=1.598\pm 0.062\,R_{\oplus}$. The phase curve of the planet is detected at a $3.7\sigma$ level with an amplitude of $10.8\pm2.9$ ppm. The planetary parameters of the outer non-transiting planet are $P_c=871.34\pm0.11$~d, $K_c=189.91\pm0.95$~m~s$^{-1}$ corresponding to a minimum mass of $M\sin{i}=5.32\pm0.20\,M_{Jup}$, and eccentricity  $e=0.842\pm0.001 $. The rotational period of the star was measured to be $P_{rot} = 18.8\pm0.4 $\,d with a decay scale of $39.2 _{-1.3}^{+2.3}$\,d. For parameters with uninformative priors, the results of this analysis perfectly agree with those presented in the previous section (see Table\,\ref{tab:juliet_pyorbit_comp} for a comparison, and Table\,\ref{tab:pyorbit_parameters} for the remaining \texttt{PyORBIT} parameters).

For completeness, we repeated the same analysis including a third planet in the model, constrained with a period between 1 day and 50\,d, that is, limited by the periastron distance of the eccentric Jupiter, using \texttt{dynesty} to compute the Bayesian evidence. With a Bayes factor of $\ln{K} = \ln{\mathcal{Z}_\textrm{2p}} - \ln{\mathcal{Z}_\textrm{3p}} = 70$, the two-planet model was strongly favored. 

\section{Discussion}\label{sec:discuss}

\subsection{HARPS-N detection sensitivity}\label{sec:sensitivity}
\begin{figure}
    \centering
    \includegraphics[width=1\linewidth]{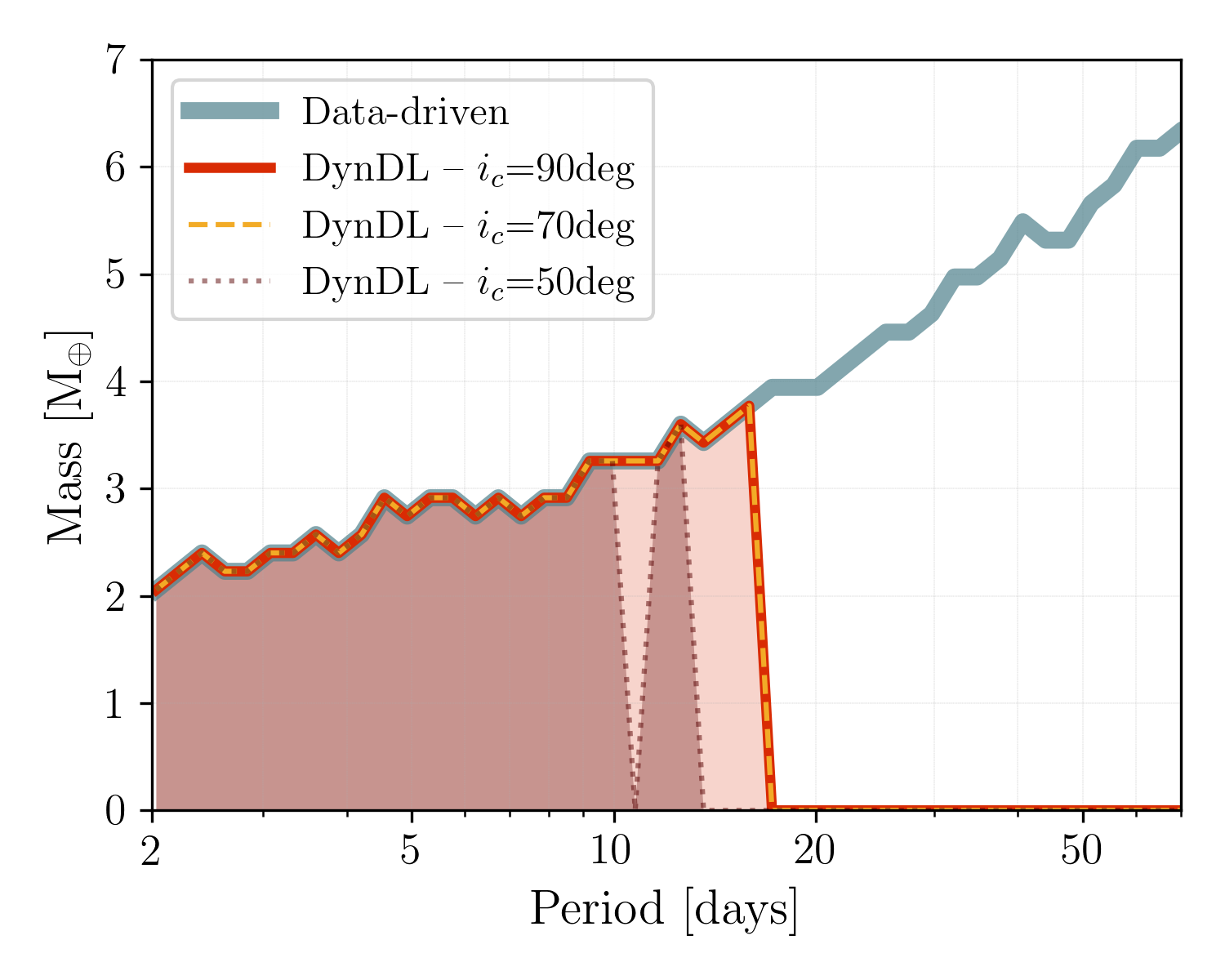}
    \caption{Planet detection limits computed with \texttt{ARDENT}. The data-driven limits are plotted in thick blue (i.e., any planet above the line can be ruled out because it would be detectable). The plain red, dashed yellow, and dotted brown lines represent the dynamical detection limits (DynDL) above which any planet is excluded either dynamically or by the data for various supposed orbital inclinations of planet c (90, 70, and 50 deg, respectively).}
    \label{fig:sensitivity}
\end{figure}

As we discuss further below, planet c has an important dynamical impact on the inner system. From the formation and evolution viewpoints, the outer giant has likely been affecting the architecture of the inner system. We computed detection limits for additional planets inner to planet c. We performed this analysis with \texttt{ARDENT}\footnote{\url{https://github.com/manustalport/ardent}} \citep{Stalport2025} to account for the dynamical effect of the outer giant. The code computes detection limits based on planet injection--recovery tests in the period--RV semi-amplitude space (data-driven detection limits). Instead of using the RV residuals, we injected signals in a white-noise time series based on the RV uncertainties, as done previously (e.g., B23). We therefore randomly generated 10 000 points in the $P$-$K$ space, following uniform distributions in [2, 100]\,d for $P$ and in [0.1, 1.2] times the RV error root mean-square (rms) for $K$. For each point, we performed ten injection--recovery tests with different evenly sampled orbital phases. The 95$\%$ detection limit curve is illustrated in blue in Fig. \ref{fig:sensitivity}, after conversion of $K$ into the planet mass. 

In a second step, \texttt{ARDENT} investigates the dynamical plausibility of injecting an additional planet in the period-mass space. The code scans the zone below the data-driven detection limits and further rejects injected planets that lead to unstable systems. The updated limits, or dynamical detection limits, combine the data and stability constraints to improve the completeness of the system. To compute these limits, \texttt{ARDENT} combines analytical and numerical stability criteria. The latter include a chaos indicator to enable a shortening of the numerical integrations. 

The orbital inclination of planet c is currently not constrained observationally. Therefore, we investigated its dynamical effect on the inner system with various inclinations and masses (the RV measurements only constrain $M\,sini$). In the \texttt{ARDENT} publication, \citet{Stalport2025} proposed a calibration of their chaos indicator based on numerical experiments with coplanar systems. Since this calibration was not tested for mutually inclined orbits, we decided to ignore the chaos indicator and instead performed brute-force long integrations over 10$^6$ years. We note that in order to ignore the chaos indicator in \texttt{ARDENT}, we set the semimajor axis drift threshold \texttt{max\_drift\_a} parameter to a high value (0.1, or 10$\%$, in this study). Effectively, no system reaches this high threshold except those that were unstable before the simulation ends (collision or close encounter). We also set the parameter \texttt{Nphases} to 8, meaning that we carried out eight stability analyses for each tested point of the period-mass space, with different evenly spaced orbital phases. The result is presented in Fig. \ref{fig:sensitivity} (cf. DynDL) with $i_c$=90, 70, and 50 degrees. 

While the results are identical between edge-on and 70 degree orbital inclinations, the dynamical effect of the outer giant is slightly larger with $i_c$=50 deg (and a mass increase in c by a factor 1.3). 
Above $\sim$10\,d, we observe secular instability. The injected planets exchange significant angular momentum with planet c, which triggers eccentricity and ultimately leads to close encounters after several tens of thousand orbits. Some of the injected planets at 11.6 and 12.6\,d present the same unstable behavior (even if nonzero, the rate of stability in these period bins is below 100$\%$). Hence, the few planets that survived will also likely experience secular instability after a longer time span. In contrast, all of the injected planets below 10\,d survived the integration. For all inclination scenarios combined, the dynamical detection limits exclude any additional planet beyond 17\,d due to the gravitational effect of the outer giant. This limit also provides strong constraints on the planet mass, as we can exclude the presence of any additional planet between b and c with a mass higher than 4 M$_{\oplus}$.

\subsection{Formation scenario}\label{sec:formation}
Its possible formation pathway makes this system particularly interesting, in particular, how K2-312\,c attained its high eccentricity. Although planet-disk interactions can drive eccentricity growth \citep{goldreich03}, it is unlikely to explain the high eccentricity of K2-312\,c \citep{duffell15}. Secular interactions, such as von Zeipel-Lidov-Kozai (ZLK) oscillations \citep{vonZeipel2910,Lidov1962,Kozai1962}, can generate high eccentricities, but require the presence of a massive external perturber. However, no evidence of stellar companions is found in high-resolution imaging (Sect.\,\ref{sec:highres}). Gaia astrometry yields a renormalized unit weight error (RUWE) of 1.1, consistent with a well-behaved single-star solution, and no comoving sources with compatible proper motions are found within $5'$ of the target.

An alternative is planet-planet scattering \citep{ford08,chatterjee08,juric08}, which can produce eccentricity distributions that largely match that of the known exoplanet population and can produce eccentricities up to $e > 0.99$ \citep{carrera19}. To consider this in more detail, we carried out an n-body analysis similar to that presented in \citet{ford08}. For simplicity, we initially did not include the USP, but assumed that the system started with two equal-mass giant planets ($M_p = 5.3$~M$_{\rm Jup}$). The initial conditions were set so that the total orbital energy matched that of K2-312\,c (i.e., $E_{\rm init} = -(M_* + M_{\rm p,c})/2 a_{\rm c}$). The initial semimajor axis of the outer of the two giant planets was taken to be $a_2 = a_1 (1 + \Delta_c)$, where $a_1$ is the initial semimajor axis of the inner of the two giant planets. The term $(1 + \Delta_c)$ is the critical semimajor axis ratio above which Hill stability is guaranteed for circular obits \citep{gladman93} with $\Delta_c = 2.4 (M_1/M_* + M_2/M_*)^{1/3}$. This gave initial semimajor axes of $a_1 = 3.12$~au and $a_2 = 4.67$~AU. The initial eccentricities and inclination were distributed uniformly in the range $0 \leq  e \leq 0.05$ and $0^\circ \leq i \leq 1^\circ$.  Initial longitudes and phases were randomly distributed between $0$ and $2 \pi$. The short temporal baseline of the K2 and TESS data prevented us from determining whether K2-312\,c is transiting or not (see Sect.\,\ref{sec:solution} for a more detailed discussion). 

Using {\sc mercury} \citep{chambers99} with the pure Bulirsch-Stoer integrator, we ran a total of 100 simulations for a time of $t = 10^7$~years. In 21 of these simulations, one of the giant planets was scattered onto an orbit just inside $2$~au with an eccentricity $e > 0.75$. The maximum eccentricity was $e = 0.87$, and in all 21 simulations, the second giant planet was ejected from the system. We then repeated the above simulations, but also included a close-in super-Earth with a mass of $M_p = 5.6$~M$_\oplus$, an initial semimajor axis of $0.05$~AU, and an initially circular orbit. We set the semimajor axis of the inner super-Earth at $a = 0.05$~au since it would need to be inside $a \sim 0.05$~au if it has to then undergo obliquity-driven tidal evolution to become a USP \citep[e.g.,][]{millholland20}. In this case, 13 of the simulations resulted in one of the giant planets being scattered onto an orbit inside $2$~au with an eccentricity $e > 0.75$. In all of these, the second giant planet was ejected and the inner super-Earth survived with a mutual inclination between the surviving eccentric giant planet and the inner super-Earth lower than $10^\circ$. We acknowledge, however, that this result is somewhat dependent on two giant planets initially. Planet-planet scattering in systems with more than two giant planets can generate higher mutual inclinations \citep{lu25}.

The above formation scenario would suggest that the USP planet and outer giant are on orbits with a relatively low mutual inclination.  Hence, the outer companion would be of planetary mass (i.e., $M_c \sin i \sim M_c$). However, the architecture of the system does not allow us to rule out a much higher mutual inclination. For a super-Earth inside $a \sim 0.05$~AU, the ZLK timescale due to perturbations from K2-312\,c is about $10^5$~years, while the general relativistic precession timescale is about $10^4$~years \citep{fabrycky07}. At its current location ($a = 0.0167$~AU), the tidal precession timescale is also likely to be substantially shorter than the ZLK timescale. This would tend to damp eccentricity growth from ZLK cycles and the inner super-Earth could survive even if the outer giant were on an orbit with a mutual inclination above $\sim 40^\circ$. 

Similarly, we cannot actually rule out a scenario in which K2-312\,c achieved its high eccentricity through ZLK cycles, driven by a more distant inclined companion, and then scattered K2-312\,b onto a high-eccentricity orbit that was then tidally circularized onto the USP orbit it now occupies. There is, however, no evidence of an additional outer companion that could drive the ZLK cycles in K2-312\,c.  On the other hand, it is intriguing that the current semimajor axis of K2-312\,b is close to where it might be expected if it had evolved through high-eccentricity migration \citep{Petrovich2019}.

The release of Gaia DR4 will allow us to determine the inclination of K2-312\,c, and, hence, indicate whether planet-planet scattering is indeed the most likely formation scenario. This in turn determines what that might imply about the origin of the USP.

\section{Conclusions}\label{sec:conclusions}

We have presented a comprehensive RV analysis of the K2-312 system, confirming the peculiar properties of the outer giant planet K2-312\,c initially reported by B23. 
Our results highlight the following key points:
\begin{enumerate}
    \item With the extended HARPS-N baseline, we confirmed the minimum mass and eccentricity of K2-312\,c to be $M_c \sin i = 5.29^{+0.20}_{-0.22}\,M_{\rm Jup}$ and $e_c = 0.8435\pm0.0013$, respectively, while we refined the measurement of the orbital period to $871.3$~d.    
    \item The coexistence of a USP super-Earth and a highly eccentric cold Jupiter allowed us to plausibly reconstruct the formation and evolution of the system, supporting the idea that violent dynamical histories involving outer giants do not necessarily destroy close-in rocky planetary systems, although they  shape their final architecture. As shown in \citet{Bonomo2025} (see their Fig.~3), the higher the CJ eccentricity, the lower the multiplicity of inner small planets.
    \item If the system started with two giant planets, then planet-planet scattering can produce the observed eccentricity of K2-312\,c while also allowing K2-312\,b to survive, as long as it was already on a reasonably close-in orbit ($a_{\rm b} \lesssim 0.05$~AU). In this wild scenario, K2-312\,c and K2-312\,b would retain a relatively low mutual inclination.
\end{enumerate}

The coming DR4 of Gaia astrometry might break the $\sin i$ degeneracy and provide the inclination and true mass of K2-312\,c. This would shed further light on the architecture and dynamics of this intriguing system.

\section{Data availability}
Table\,\ref{tab:dataset} is only available in electronic form at the CDS via anonymous ftp to cdsarc.u-strasbg.fr (130.79.128.5) or via http://cdsweb.u-strasbg.fr/cgi-bin/qcat?J/A+A/.

\begin{acknowledgements}
This work is based on observations made with the Italian Telescopio Nazionale {\it Galileo} (TNG) operated by the Fundaci\'on Galileo Galilei (FGG) of the Istituto Nazionale di Astrofisica (INAF) at the Observatorio del Roque de los Muchachos (La Palma, Canary Islands, Spain).  The observations from March to June 2020 were performed in full service mode due to the COVID pandemic. The authors gratefully acknowledge the efforts made by the TNG staff on that occasion. The HARPS-N project was funded by the Prodex Program of the Swiss Space Office (SSO), the Harvard University Origin of Life Initiative (HUOLI), the Scottish Universities Physics Alliance (SUPA), the University of Geneva, the Smithsonian Astrophysical Observatory (SAO), the Italian National Astrophysical Institute (INAF), University of St. Andrews, Queen’s University Belfast, and University of Edinburgh. L.N. acknowledges financial contribution from the INAF Large Grant 2023 ``EXODEMO''. L.P acknowledges funding from the Royal Society Career Development Fellowship, grant number CDF R1 251054. F.P.E. would like to acknowledge the Swiss National Science Foundation (SNSF) for supporting research with HARPS-N through the SNSF grants nr. 140649, 152721, 166227, 184618 and 215190. Y.N.E.E. acknowledges support from a Science and Technology Facilities Council (STFC) studentship, grant number ST/Y509693/1. The HARPS-N Instrument Project was partially funded through the Swiss ESA-PRODEX Programme. MLM is supported by individual research time under NASA contracts NAS5-26555 and NAS5-03127 to the Associated Universities for Research in Astronomy for the operation of the Hubble Space Telescope and the James Webb Telescope Science Operations Centers at STScI. AM acknowledges a UK Science and Technology Facilities Council (STFC) small grant ST/Y002334/1 and funding from a UKRI Future Leader Fellowship, grant number MR/X033244/1.
This work has made use of the Python packages \texttt{numpy}, \texttt{pandas}, \texttt{matplotlib}, \texttt{PyORBIT}, \texttt{S+LEAF}, \texttt{pyDE}, \texttt{lightkurve}, \texttt{dynesty}, \texttt{batman}, \texttt{astroquery}, \texttt{juliet} and packages therein.
\end{acknowledgements}

%-------------------------------------------------------------------
\bibliographystyle{aa}
\bibliography{aa61259-26}

\begin{appendix}

\section{Additional figures and tables}

\begin{figure}[!h]
\centering
\includegraphics[width=1\linewidth]{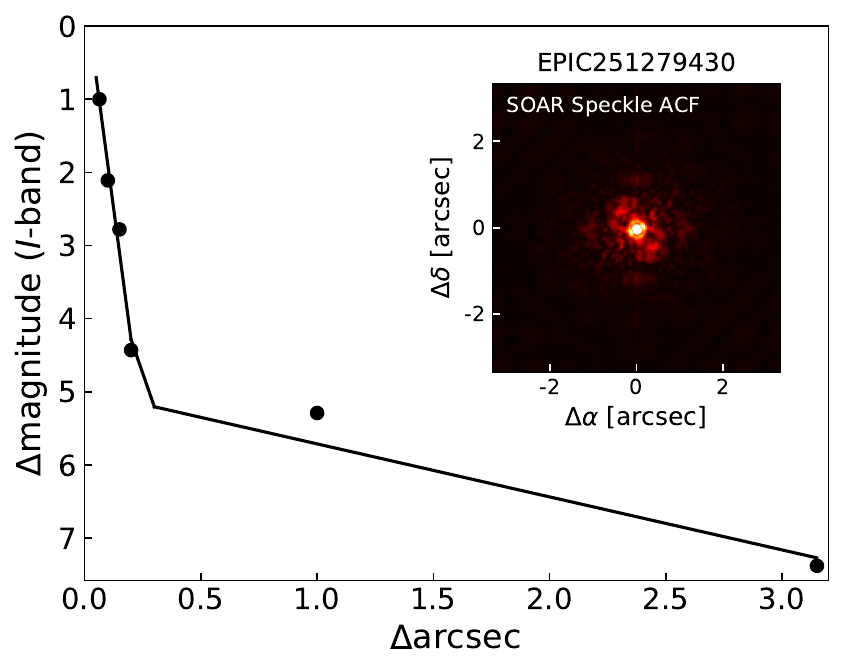}
\caption{K2-312 detection sensitivity to nearby sources from the SOAR telescope (Sect.\,\ref{sec:highres}).} 
\label{fig:highres}
\end{figure}

\begin{figure}[!h]
\centering
\includegraphics[width=1\linewidth]{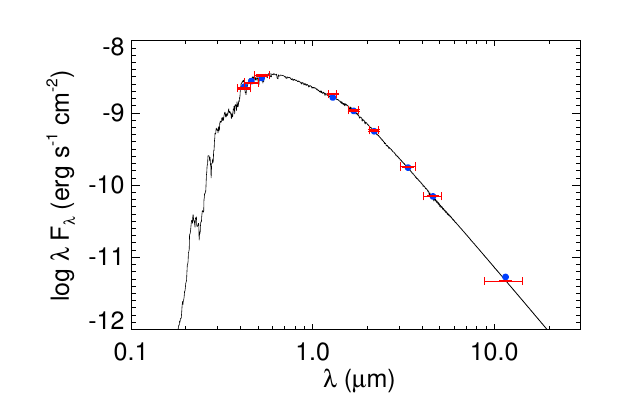}
\caption{Stellar spectral energy distribution. The broadband measurements converted from the Tycho-2 $B_{\rm T}$ and $V_{\rm T}$, APASS Johnson $B$, 2MASS $J$, $H$, and $K_{\rm s}$, and WISE $W1$, $W2$, and $W3$ magnitudes are shown in red, and the corresponding theoretical values with blue circles. The solid black line displays the non-averaged best-fit model.} 
\label{fig:stellarSED}
\end{figure}

\begin{figure}
    \centering
    \includegraphics[width=1\linewidth]{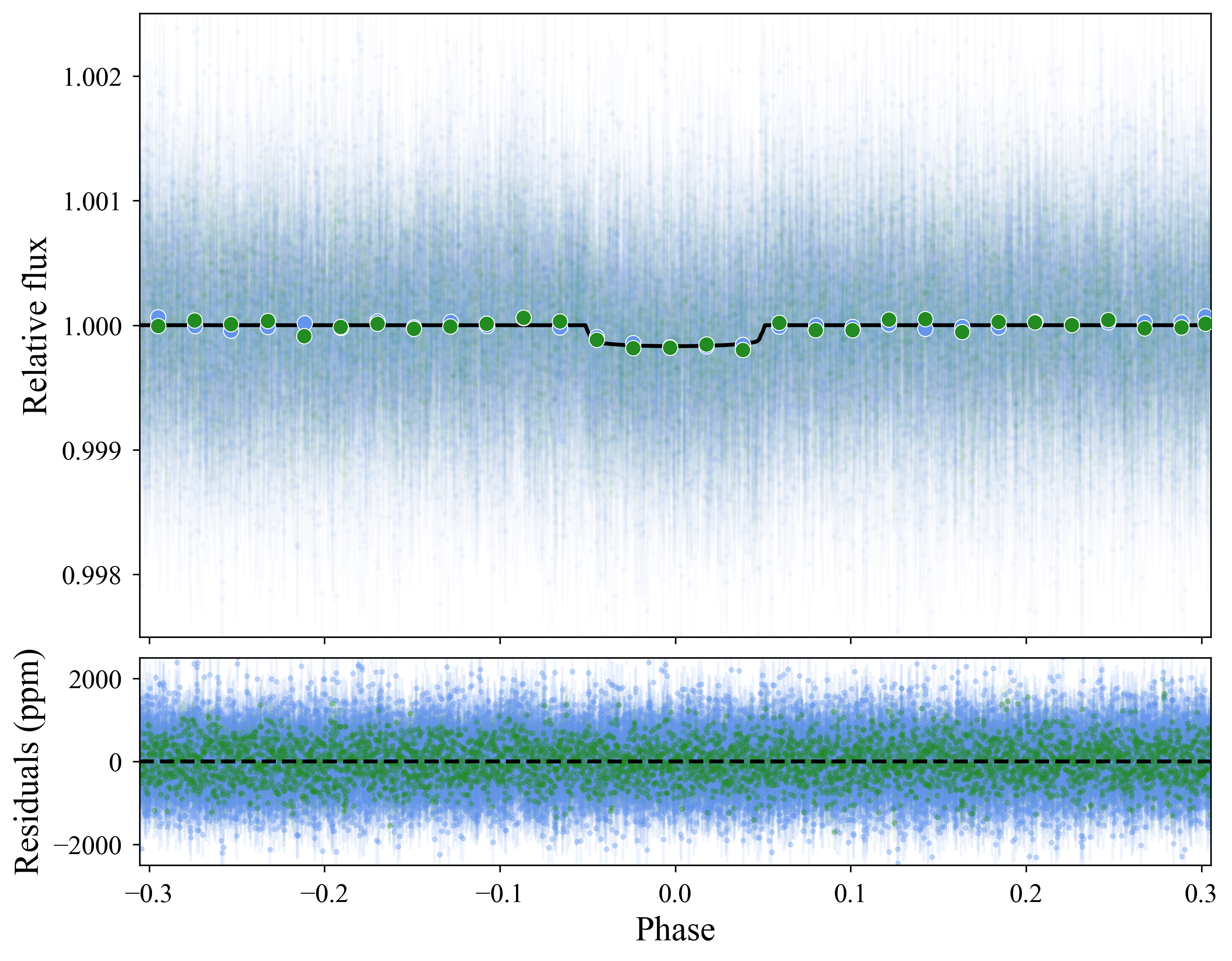}
    \caption{Zoomed-out version of the right panel of Fig.\,\ref{fig:transit} (i.e. the phase-folded TESS transits).}
    \label{fig:zoom}
\end{figure}

\begin{figure*}[!h]
    \centering
    \includegraphics[width=0.65\linewidth]{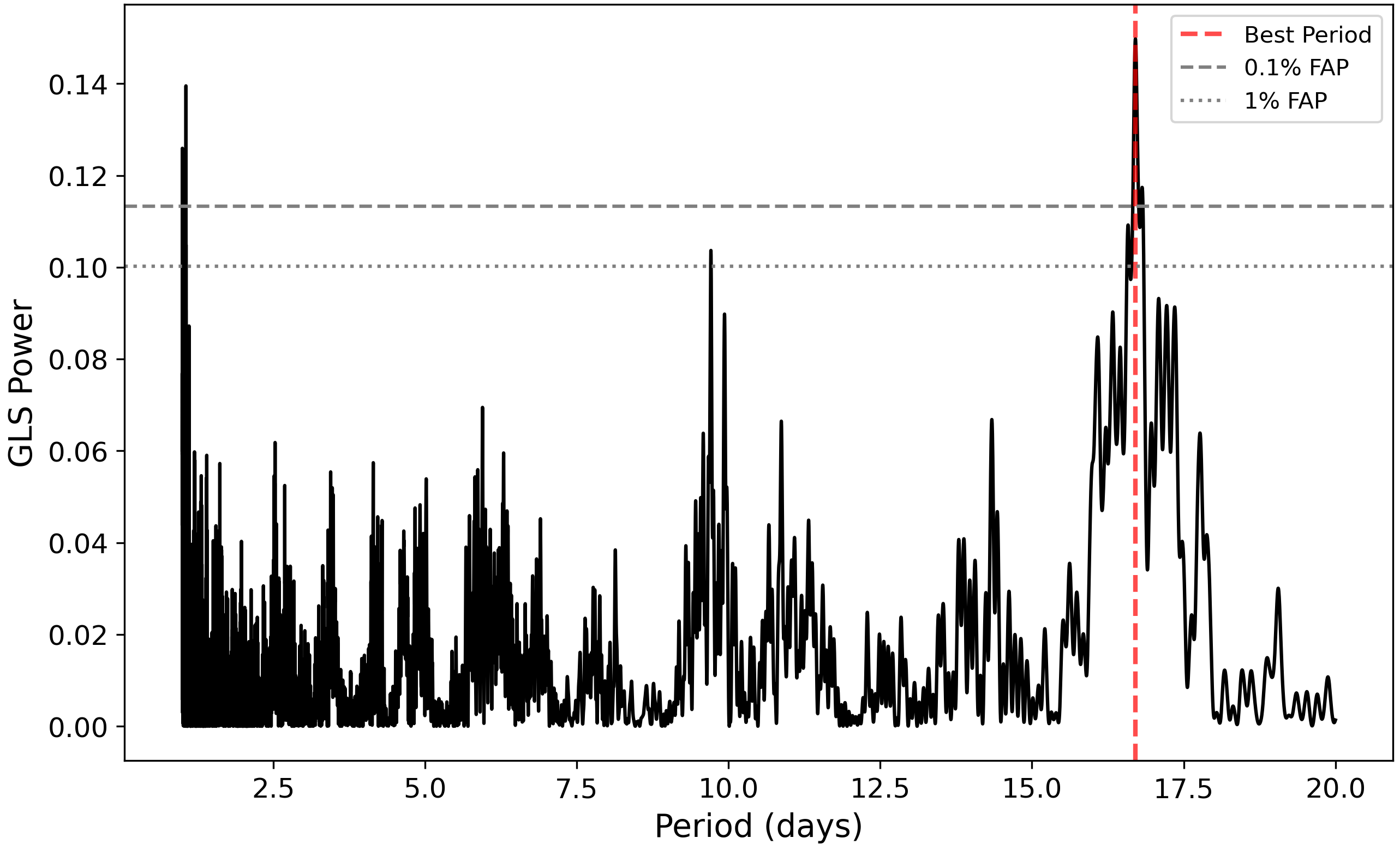}
    \caption{GLS periodogram of the RV residuals from the 2-planet model plus a trend model.}
    \label{fig:GLS_res}
\end{figure*}

%\clearpage

\begin{figure*}
    \centering
    \includegraphics[width=0.80\linewidth]{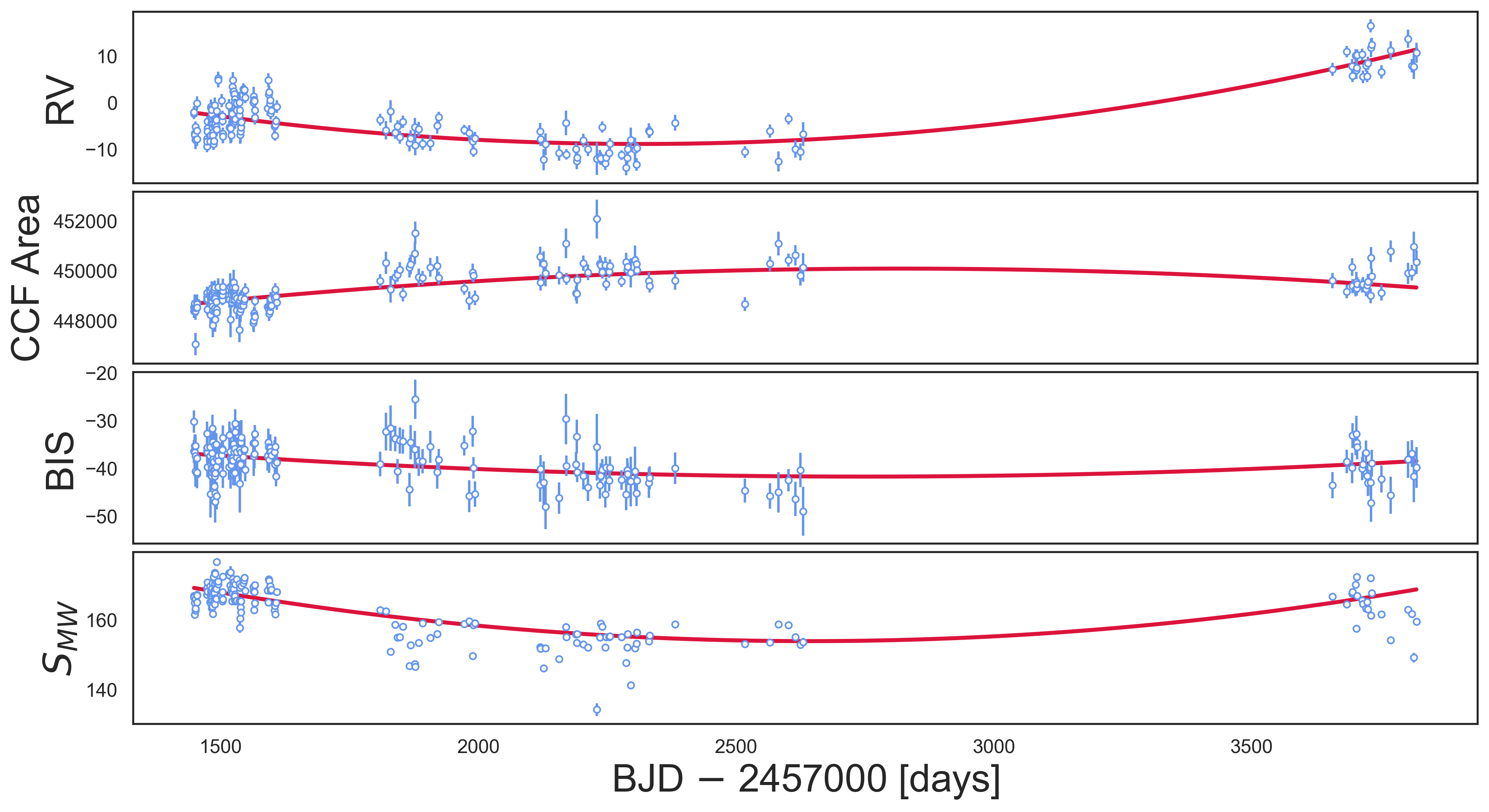}
    \caption{RV residuals after subtraction of the planetary models (top panel), together with the CCF area, BIS, and Mount-Wilson S-index (lower panels) derived from the HARPS-N spectra. The red line represents a simple quadratic fit of the data. Some spectra have been excluded because they show strong outliers in at least one activity indicator.}
    \label{fig:trend}
\end{figure*}

\begin{figure*}
    \centering
    \includegraphics[width=0.80\linewidth]{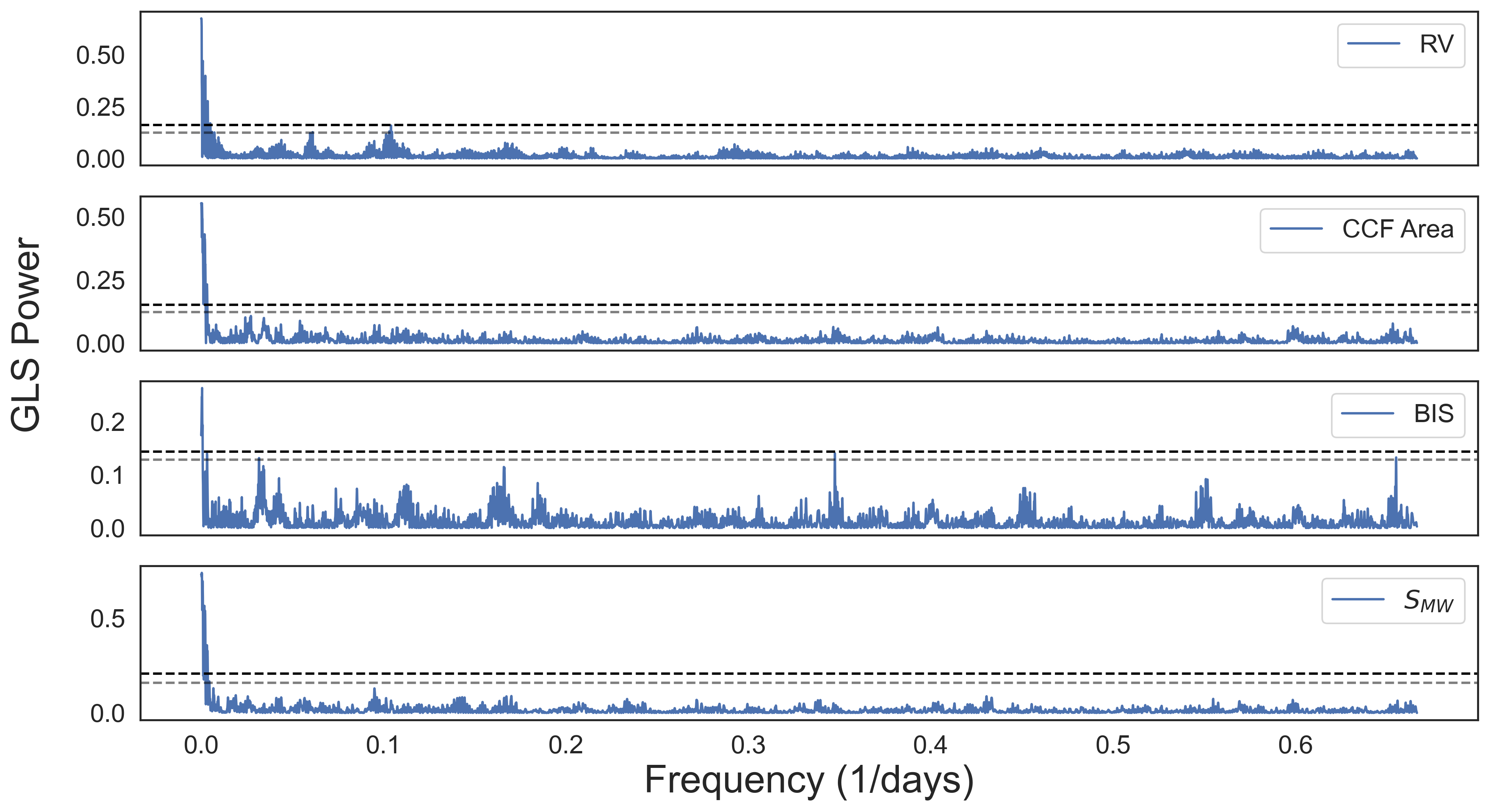}
    \caption{GLS periodogram of the datasets displayed in Fig.\,\ref{fig:trend}. The horizontal dashed lines mark the 1\% and 0.1\% FAP, evaluated with the bootstrap method.}
    \label{fig:GLS}
\end{figure*}

\begin{table*}
\centering
\caption[]{K2-312 HARPS-N RV data points and activity indices.}\label{tab:dataset}
{\tiny\renewcommand{\arraystretch}{.8}
\resizebox{!}{.05\paperheight}{%
\begin{tabular}{llllllllllll}
    \hline\hline \\[-6pt]
    $\mathrm{BJD_{\textsf{TDB}}}$ & RV & $\pm1\upsigma_{\textsf{RV}}$ & FWHM & $\upsigma_{\textsf{FWHM}}$ & BIS & $\upsigma_{\textsf{BIS}}$ & Contrast & $\upsigma_{\textsf{Cont}}$ & $S_{MW}$ & $\upsigma_{\mathrm{S_{MW}}}$ \\ %& $\log R^{\prime}_{\rm HK}$ & $\upsigma_{\log R^{\prime}_{\rm HK}}$ \\ %& $I_{\rm Ca~{\sc ii}}$ & $I_{\rm H\upalpha06}$ & $I_{\rm He~{\sc i}}$ & $I_{\rm Na~{\sc i}}$ & $I_{\rm Ca~{\sc i}}$ & $I_{\rm H\upalpha16}$ \\ 
    $-2400000\,[d]$ & $\mathrm{~~[m\,s^{-1}]}$ & $\mathrm{~~[m\,s^{-1}]}$ & & & & & & & & & \rule[-0.8ex]{0pt}{0pt} \\ 
    \hline \\
    58448.64706 & 8296.569 & 1.166499 & 8050.332 & 2.332997 & -30.2159 & 2.332997 & 55.71648 & 0.016147 & 166.9017 & 0.287259 \\
    58448.77348 & 8296.572 & 1.231099 & 8051.073 & 2.462197 & -36.4656 & 2.462197 & 55.69195 & 0.017032 & 166.4503 & 0.315335 \\
    58449.67051 & 8287.274 & 1.633332 & 8055.822 & 3.266663 & -36.6805 & 3.266663 & 55.69759 & 0.022586 & 161.4141 & 0.513544 \\
    58449.77392 & 8285.442 & 1.209178 & 8044.682 & 2.418356 & -35.3491 & 2.418356 & 55.74693 & 0.016758 & 164.895  & 0.300624 \\
    58451.62921 & 8291.458 & 1.626734 & 8045.016 & 3.253468 & -40.6209 & 3.253468 & 55.73551 & 0.02254  & 162.6831 & 0.518821 \\
    58451.71566 & 8292.300   & 1.166524 & 8042.324 & 2.333049 & -35.3194 & 2.333049 & 55.76809 & 0.016178 & 163.2008 & 0.275843 \\
    \dotfill & \dotfill & \dotfill & \dotfill & \dotfill & \dotfill & \dotfill & \dotfill & \dotfill & \dotfill & \dotfill \rule[-0.8ex]{0pt}{0pt} \\
    \bottomrule
\end{tabular}}}
\tablefoot{The full Table is available at the CDS.}
\end{table*}

\begin{table*}
\centering
\caption{Comparison between the \textrm{juliet} and \textrm{PyORBIT} analysis.}\label{tab:juliet_pyorbit_comp}
\resizebox{0.85\hsize}{!}{
\begin{tabular}{lcccc}
    \hline\hline
     & \multicolumn{2}{c}{\textrm{juliet}} & \multicolumn{2}{|c}{\textrm{PyORBIT}} \\
    \hline
    Parameters & Prior & Fitted & Prior & Fitted \\
\hline \\[-6pt]
\multicolumn{5}{l}{\large{\,\,\,K2-312 b}} \\[2pt]
$K$ (m\,s$^{-1}$)\dotfill & $\mathcal{U}[0, 10]$ & $3.66\pm0.20$ & $\mathcal{U}[0, 20]$ & $3.61 \pm 0.22$ \\
$P_{\rm orb}$ (d)\dotfill & $\mathcal{U}[0.719, 0.720]$ & $0.71957926^{+0.00000042}_{-0.00000036}$ & $\mathcal{U}[0.719, 0.720]$ & $0.71957983_{-0.00000060}^{+0.00000065}$  \\
$T_{\rm 0}$ (BJD - $2\,458\,000$)\dotfill & $\mathcal{U}[95, 96]$ & $95.56741^{+0.00051}_{-0.00052}$ & $\mathcal{U}[95, 96]$ &  $95.56636_{-0.00056}^{+0.00054}$ \\
$R_{\rm p}/R_{\star}$\dotfill & $\mathcal{U}[0, 0.1]$ & $0.01242^{+0.00022}_{-0.00029}$ & $\mathcal{U}[0, 0.1]$ & $0.01228_{-0.00039}^{+0.00028}$ \\ 
$b$\dotfill & $\mathcal{U}[0, 1]$ & $0.429^{+0.075}_{-0.107}$ & $\mathcal{U}[0, 0.8]$ & $0.396_{-0.15}^{+0.094}$  \\
$i$ (deg) [Derived]\dotfill & & $81.7^{+2.2}_{-1.8}$ & &  $82.4_{-2.1}^{+2.9}$   \\ 
$\rho_*$\tablefootmark{a} (g\,cm$^{-3}$, $\rho_\odot$) & $\mathcal{N}(0.941, 0.095)$ & $0.960^{+0.096}_{-0.094}$ & $\mathcal{N}(0.667, 0.067)$ &  $0.683 \pm 0.067 $ \\
$\delta_{\rm ecl}$ (ppm)\dotfill & $\mathcal{U}[0, 500]$ & $9.1^{+2.4}_{-2.7}$ & $\mathcal{U}[0, 1]$ & $10.8 \pm 2.9$ \\ [2pt]

\multicolumn{5}{l}{\large{\,\,\,K2-312 c}} \\[2pt]
$K$ (m\,s$^{-1}$)\dotfill & $\mathcal{U}[0, 300]$ & $191.6\pm1.3$ & $\mathcal{L}_2[0.01, 1000]$ & $189.91 \pm 0.095$ \\
$P_{\rm orb}$ (d)\dotfill & $\mathcal{U}[750, 1200]$ & $871.32\pm0.13$ & $\mathcal{L}_2[800, 1000]$ & $871.34 \pm0.11$ \\
$\lambda$\tablefootmark{b} (deg) \dotfill & $\mathcal{U}[0, 360]$ & $77.2\pm0.5$ & $\mathcal{U}[0, 360]$ & $78.0\pm0.3$ \\
$\sqrt{e}\sin\omega_p$\dotfill & $\mathcal{U}[-1, 1]$ & $0.6060\pm0.0056$ & $\mathcal{U}[-1, 1]$ & $0.6141\pm 0.0038$\\
$\sqrt{e}\cos\omega_p$\dotfill & $\mathcal{U}[-1, 1]$ & $0.6901^{+0.0050}_{-0.0049}$ & $\mathcal{U}[-1, 1]$ &  $0.6818\pm0.0035$ \\
$e$ [Derived]\dotfill &  & $0.8435\pm0.0013$ &  & $0.8420 \pm 0.0009 $ \\ 
$\omega_p$ (deg) [Derived]\dotfill & & $41.3\pm0.5$ &  & $42.01_{-0.31}^{+0.32}$  \\  [2pt]

\multicolumn{5}{l}{\large{\,\,\,Instrumental}} \\[2pt]
$q_1$ (K2)\dotfill & $\mathcal{U}[0, 1]$ & $0.17^{+0.20}_{-0.10}$ & $\mathcal{U}[0, 1]$ & $0.30_{-0.19}^{+0.37}$ \\
$q_2$ (K2)\dotfill & $\mathcal{U}[0, 1]$ & $0.38^{+0.37}_{-0.26}$ & $\mathcal{U}[0, 1]$ & $0.31_{-0.21}^{+0.36}$ \\
$q_1$ (TESS)\dotfill & $\mathcal{U}[0, 1]$ & $0.15^{+0.26}_{-0.11}$ & $\mathcal{U}[0, 1]$ & $0.22_{-0.16}^{+0.33}$ \\
$q_2$ (TESS)\dotfill & $\mathcal{U}[0, 1]$ & $0.34^{+0.38}_{-0.25}$ & $\mathcal{U}[0, 1]$ & $0.39_{-0.28}^{+0.36}$ \\ [2pt]

\multicolumn{5}{l}{\large{\,\,\,GP, K2}} \\[2pt]
$\sigma_{\textsf{K2}}$ (ppt)\dotfill & $\mathcal{LU}[0, 1000]$ & $34.8\pm1.5$ & $\mathcal{LU}[0, 1000]$ & $24.6\pm1.2$ \\
$P_{\mathrm{rot}}$ (d)\dotfill & & & $\mathcal{U}(2,22 )$  & $5.437_{-0.063}^{+0.052}$ \\
$P_{\mathrm{dec}}$ (d)\dotfill & & & $P_{\mathrm{dec}} > 2 P_{\mathrm{rot}}$ & $10.96 \pm 0.15$ \\
$O_{\mathrm{amp}}$ \dotfill & & & $\mathcal{N}(0.350, 0.035)$ &  $0.471 \pm 0.024$\\
$H_{\mathrm{rot}}$ (ppm)\dotfill & $\mathcal{U}(0, 10^5)$ & $592\pm25$ & $\mathcal{U}(0, 5 \cdot 10^4)$ & $1860_{-140}^{+160}$ \\ [2pt]

\multicolumn{5}{l}{\large{\,\,\,GP, HARPS-N}} \\[2pt]
$\overline{\mu}_{\textsf{HARPS-N}}$ (m\,s$^{-1}$)\dotfill & $\mathcal{U}[8300, 8350]$ & $8330\pm4$ & $\mathcal{U}[8000, 8500]$ & $8321.4 \pm 0.5 $ \\
$\sigma_{\textsf{HARPS-N}}$ (m\,s$^{-1}$)\dotfill & $\mathcal{U}[0, 10]$ & $0.80^{+0.24}_{-0.19}$ & $\mathcal{U}[0, 20]$ & $1.45_{-0.33}^{+0.31}$ \\
$P_{\mathrm{rot}}$ (d)\dotfill & $\mathcal{U}[15, 25]$ & $18.98^{+0.66}_{-0.74}$ & $\mathcal{U}[10, 22]$ & $ 18.80_{-0.32}^{+0.39}$ \\
$P_{\mathrm{dec}}$ (d)\dotfill & $P_{\mathrm{dec}} >  P_{\mathrm{rot}}/2$ & $13.78^{+3.11}_{-2.73}$ & $P_{\mathrm{dec}} > 2 P_{\mathrm{rot}}$ &  $39.2_{-1.3}^{+2.3}$ \\
$O_{\mathrm{amp}}$ \dotfill & $\mathcal{U}[0, 1]$ & $0.062^{+0.019}_{-0.009}$ & $\mathcal{N}(0.350, 0.035)$ &  $0.373 \pm  0.029$ \\
$V_{\mathrm{con}}$\tablefootmark{c} (m\,s$^{-1}$)\dotfill & $V_{\mathrm{con}} > 0$ & $3.0^{+0.29}_{-0.26}$ & $V_{\mathrm{con}} > 0$ & $2.70 \pm 0.40$ \\
$V_{\mathrm{rot}}$\tablefootmark{d} (m\,s$^{-1}$)\dotfill & & & $\mathcal{U}[-50, 50]$ & $3.6 \pm 1.0$  \\
\hline
\end{tabular}
}
\tablefoot{Best-fit median values, with upper and lower 68\% credibility bands as errors, as extracted from the posterior distribution of the relative models. Priors are uniform in natural space unless specified. We used an MCMC sampler for the \texttt{PyORBIT} analysis; hence, the different boundaries compared to \texttt{juliet} had no effect on the results.     
 $\mathcal{L}_2$: parameters explored in Logarithmic base 2 space. $\mathcal{N}$: Normal prior. $\mathcal{U}$: uniform prior. $\mathcal{LU}$: Log-uniform prior. \\
 \tablefoottext{a}{\texttt{juliet} and \texttt{PyORBIT} use different units for the density (i.e. g\,cm$^{-3}$ and $\rho_\odot$).}
 \tablefoottext{b}{Mean longitude at epoch $T_{\rm ref} = 2456000.0$, assuming a longitude of the ascending node $\Omega = 180 ^{\circ}$.}
 \tablefoottext{c}{Corresponding to $H_{\mathrm{amp}}$ when the first derivative of the GP is not employed.}
 \tablefoottext{d}{Amplitude of the first derivative of the GP.}
 }
\end{table*}

%\clearpage

\begin{table*}
\centering
\caption{Additional model parameters for the \textrm{PyORBIT} analysis.}\label{tab:pyorbit_parameters}
\renewcommand{\arraystretch}{1.2}
\resizebox{1.0\hsize}{!}{
\begin{tabular}{lccccccc}
    \hline\hline
     & K2 & TESS & RV & BIS & FWHM & $S_{MW}$ & H$\alpha$ \\
     & (unit) & (unit) & (m/s) & (m/s) & (m/s) &  &  \\
\hline \\[-6pt]

$\sigma_{\rm w}$ & $24.6 \pm 1.2 \cdot 10^{-6}$ &  $351.4 \pm 5.1 \cdot 10^{-6}$ & $ 1.46 \pm 0.22 $ & $ 1.41_{-0.44}^{+0.37}$ &   $4.59 \pm 0.36$ &  $3.85 \pm 0.22$ & $2.83 \pm 0.14$ \\
$\overline{\mu}$\tablefootmark{a} & $1.00175 \pm 0.00090$ & & $8321.37 \pm 0.60$ & $ -39.37 \pm 0.37$ & $8042.62 \pm 0.93$ & $ 157.93 \pm 0.53 $ & $194.98 \pm 0.29 $ \\
$V_{\mathrm{con}}$\tablefootmark{b} & $1860 _{-140}^{+160} \cdot 10^{-6} $ & & $2.70_{-0.36}^{+0.40}$ & $ 0.99 \pm 0.29 $ & $3.67_{-0.57}^{+0.64}$ & $ 1.76_{-0.38}^{+0.43} $ & $ 0.28 \pm 0.24 $ \\
$V_{\mathrm{rot}}$ & & & $3.60 \pm 1.00$ & $-2.27 \pm 0.77$ & 0 (fixed) & 0 (fixed) & 0 (fixed) \\
$x_0$ (BJD-$2\,458\,000$)\tablefootmark{c} & 134.8430 & & 966.59 & 971.2092 &  964.9216 & 964.9216 & 964.9216 \\
$c_1$ ${\rm u}/s \cdot 10^{-3}$ & $0.055 \pm 0.022$ & & $-7.3 \pm 1.6 $ & $-3.88_{-0.89}^{+0.85} $ &  $-19.8 \pm 2.4 $ &  $ -16.4 \pm 1.3 $ & $ -4.54 \pm 0.62 $ \\
$c_2$ ${\rm u}/s^2 \cdot 10^{-6}$ & $1.46 \pm 0.81$ & & $9.7 \pm 1.1 $ & $1.86 \pm 0.62 $ & $9.4 \pm 1.6 $ &  $10.93 \pm 0.86 $ & $ 3.40 \pm 0.45 $ \\
\hline
\end{tabular}
}
\tablefoot{Best-fit median values, with upper and lower 68\% credibility bands as errors, as extracted from the posterior distribution. $c_1$ and $c_2$ are the coefficients of the polynomial trend, and the zero-order coefficient is absorbed by the offset parameter. \\
\tablefoottext{a}{Corresponding to the normalisation factor for the photometric dataset.}
 \tablefoottext{b}{Corresponding to $H_{\mathrm{amp}}$ when the first derivative of the GP is not employed}
\tablefoottext{c}{Offset for the independent variable, automatically fixed by the code at runtime.}
 }
\end{table*}

\end{appendix}
\end{document}